\documentclass[12pt]{article}
\usepackage{amsmath,amssymb,bbold,epsfig}
\usepackage{amsfonts}
\usepackage{bbm}
\usepackage[colorlinks=true,linkcolor=red,citecolor=green,urlcolor=cyan,
bookmarks=true,bookmarksopen=true,pdfpagemode=None,pdfstartview=FitH]{hyperref}
\newcommand{\appsection}{\addtocounter{section}{1}\setcounter{equation}{0}
                         \renewcommand{\thesection}{\Alph{section}}
}
\renewcommand{\theequation}{\arabic{equation}}
\newcommand{\be}{\begin{equation}}
\newcommand{\ee}{\end{equation}}
\newcommand{\bea}{\begin{eqnarray}}
\newcommand{\eea}{\end{eqnarray}}
\begin{document}

\title{
\vglue -0.3cm
\vskip 0.5cm
\Large \bf
Paradoxes of neutrino oscillations}
\author{
{E. Kh. Akhmedov$^{a,b}$\thanks{email: \tt 
akhmedov@mpi-hd.mpg.de}~~\,and
\vspace*{0.15cm} ~A. Yu. Smirnov$^{c,d}$\thanks{email:
\tt smirnov@ictp.it}
} \\
{\normalsize\em $^a$Max--Planck--Institut f\"ur Kernphysik,
Postfach 103980} \\ {\normalsize\em D--69029 Heidelberg, Germany
\vspace*{0.15cm}}
\\
{\normalsize\em $^{b}$National Research Centre Kurchatov
\vspace*{-0.1cm}Institute}\\{\normalsize\em Moscow, Russia 
\vspace*{0.15cm}}
\\
{\normalsize\em $^{c}$The Abdus Salam International Centre for Theoretical    
Physics} \\
{\normalsize\em Strada Costiera 11, I-34014 Trieste, Italy 
\vspace*{0.15cm}}
\\
{\normalsize\em $^{d}$Institute for Nuclear Research, Russian Academy of
Sciences} \\ {\normalsize\em Moscow, Russia}
}
\maketitle
\thispagestyle{empty}
\vspace{-0.8cm}
\begin{abstract}
\noindent
Despite the theory of neutrino oscillations being rather old, some of its 
basic issues are still being debated in the literature. We discuss,
in the framework of the wave packet approach, 
a number of such issues, including the relevance of the ``same energy'' 
and ``same momentum'' assumptions, the role of quantum-mechanical 
uncertainty relations in neutrino oscillations, the dependence of the 
production/detection and propagation coherence conditions that ensure the 
observability of neutrino oscillations on neutrino energy and momentum 
uncertainties,  the question of (in)dependence of the oscillation 
probabilities on the neutrino production and detection processes, the 
applicability limits of the stationary source approximation, and Lorentz 
invariance of the oscillation probability. We also develop a novel approach 
to calculation of the oscillation probability in the wave packet picture, 
based on the summation/integration conventions different from the standard 
one, which gives a new insight into the oscillation phenomenology. We discuss 
a number of apparently paradoxical features of the theory of neutrino 
oscillations. 
\end{abstract}

\vspace{1.cm}
\vspace{.3cm}

\newpage


\section{\label{sec:intro}Introduction}

More than 50 years have already passed since the idea of neutrino oscillations 
was put forward \cite{Pont,MNS}, and over 10 years have passed since the 
experimental discovery of  this phenomenon \cite{disc}. However, surprisingly 
enough, a number of basic issues of the theory of neutrino oscillations are 
still being debated. Moreover, some features of the theory appear rather 
paradoxical. The issues that are still under discussion include

\begin{itemize}
\item[(1)] 
Why do the often used same energy and same momentum assumptions for neutrino 
mass eigenstates composing a given flavour state, which are known to be 
both wrong, lead to the correct result for the oscillation probability?

\item[(2)]
What is the role of quantum-mechanical uncertainty relations in neutrino 
oscillations? 

\item[(3)]
What determines the size of the neutrino wave packets?

\item[(4)]
How do the neutrino production/detection and propagation coherence conditions 
that ensure the observability of neutrino oscillations depend on neutrino 
energy and momentum uncertainties?

\item[(5)]
Are wave packets actually necessary for a consistent description of neutrino 
oscillations?

\item[(6)]
When can the oscillations be described by a universal (i.e., production and 
detection process independent) probability? 

\item[(7)]
When is the stationary source approximation valid?

\item[(8)]
Would  recoillessly emitted and absorbed neutrinos (produced and 
detected in M\"ossbauer-type experiments) oscillate?  

\item[(9)]
Are oscillations of charged leptons possible? 
\end{itemize}

In the present paper we consider the first seven issues listed above, trying 
to look at them from different perspectives. We hope that our discussion will 
help clarify these points. For the last two issues, we refer the reader to 
the recent discussions in \cite{AKL1,Bil1,AKL2} (for oscillations of 
M\"ossbauer neutrinos) and \cite{Akh2} (for oscillations of charged leptons). 
We also discuss Lorentz invariance of the oscillation probability in
the wave packet picture of neutrino oscillations. 

The structure of the paper is as follows. In Sec.~\ref{sec:same} we review 
the standard derivations of the neutrino oscillation probability in the plane 
wave and stationary state approaches and point out their inconsistencies. In 
Sec.~\ref{sec:why} we introduce a general wave packet formalism for neutrino 
oscillations which does not rely on the specific form of the wave packets and 
discuss, on its basis, why the ``same energy'' and ``same momentum'' 
approaches give the correct oscillation probability. The reader not interested 
in the latter question can skip Sec.~\ref{sec:phase}. In Sec.~\ref{sec:another} 
we develop a novel approach to calculation of the oscillation 
probability in the wave packet picture, based on the alternative 
summation/integration conventions.  Sec.~\ref{sec:QM} is devoted to the 
discussion of the role of the quantum-mechanical uncertainty relations 
in neutrino oscillations. Here we consider, in particular, the question 
of what determines the size of the neutrino wave packets, as well as the 
conditions for coherent neutrino production and detection and the issues 
of coherence loss on the way between the neutrino source and detector 
and its possible restoration at detection. In Sec.~\ref{sec:stat} we 
consider the stationary source approximation for neutrino oscillations; 
we derive this approximation in the wave packet picture and analyze its 
domain of applicability. In Sec.~\ref{sec:when} we discuss the question 
of when the oscillation probability is independent of the neutrino 
production and detection processes. Sec.~\ref{sec:Lorentz} is devoted to 
the discussion of the Lorentz invariance of the oscillation probability. 
We add a few comments on the wave packet approach in 
Sec.~\ref{sec:remarks} and summarize our results in Sec.~\ref{sec:disc}. 
Appendices A and B contain some auxiliary material. 
Sections \ref{sec:another}, \ref{sec:stat}, \ref{sec:when} and 
\ref{sec:Lorentz} are largely independent of each other and can be read 
separately. 

We plan to update this article in the future if and when new 
unresolved questions or paradoxes of the theory of 
neutrino oscillations come to light.

\section{\label{sec:same}Same energy or same momentum?} 

In most derivations of the so-called standard formula for the 
probability of neutrino oscillations in vacuum (see eq.~(\ref{eq:P2}) below), 
usually the assumptions that the neutrino mass eigenstates composing a given 
flavour eigenstate either have the same momentum \cite{Eliezer:1975ja,
Fritzsch:1975rz,Bilenky:1976yj,Bilenky:1978nj} or the same energy 
\cite{Lipkin:1995cb,Grossman:1996eh,Stodolsky:1998tc,Lipkin:2000mz} are made. 
The derivation typically proceeds as follows. 

First, recall that in the basis in which the mass matrix of charged leptons 
has been diagonalized the fields describing the massive neutrinos $\nu_i$ 
and flavour-eigenstate neutrinos $\nu_a$ and the corresponding states 
$|\nu_i\rangle$ and $|\nu_a\rangle$ are related by 
\be
\nu_a~=~\sum_i U_{ai}\,\nu_i\,, \qquad \qquad
|\nu_a\rangle~=~\sum_i U_{ai}^*\,|\nu_i\rangle\,,
\label{eq:mix}
\ee
where $U$ is the leptonic mixing matrix. If one now assumes that all the mass 
eigenstates composing the initially produced flavour state $|\nu(0)\rangle
= |\nu_a\rangle$ have the same momentum, then, after time $t$ has elapsed, the 
$i$th mass eigenstate will simply pick up the phase factor $\exp(-i E_i t)$, 
and the evolved state $|\nu(t)\rangle$ will be given by
\be
|\nu(t)\rangle~=~\sum_i U_{ai}^*\,e^{-i E_i t}|\nu_i\rangle\,.
\label{eq:evol1} 
\ee 
Projecting this state onto the flavour state $|\nu_b\rangle$ and taking the 
squared modulus of the resulting transition amplitude, one gets the 
probability of the neutrino flavour transition $\nu_a\to \nu_b$ after the 
time interval $t$: 
\be 
P(\nu_a\to\nu_b; t) = \Big|\sum_i U_{bi}\; e^{-i E_i t} \;U_{ai}^*\Big|^2\,. 
\label{eq:P1} 
\ee 
Next, taking into account that the energy $E_i$ of a relativistic 
neutrino of mass $m_i$ and momentum $\bf p$ is 
\be 
E_i~=~\sqrt{p^2+m_i^2}~\simeq~p+\frac{m_i^2}{2p}\,,
\label{eq:emr1} 
\ee 
and that for relativistic pointlike particles the distance $L$ they propagate 
during the time interval $t$ satisfies 
\be 
L~\simeq~t\,, 
\label{eq:Lt} 
\ee 
one finally finds 
\be 
P(\nu_a\to\nu_b; L) = \Big|\sum_i U_{bi}\; e^{-i \frac{\Delta m_{ij}^2}{2p} L} 
\;U_{ai}^*\Big|^2 \,, 
\label{eq:P2} 
\ee 
where $\Delta m_{ij}^2=m_i^2-m_j^2$ and the index $j$ corresponds to any of 
the mass eigenstates. This is the standard formula describing neutrino 
oscillations in vacuum.%
\footnote{Let us stress that in this paper by the standard oscillation 
probability we will always understand the non-averaged (i.e., $p$- and 
$L$-dependent) probability (\ref{eq:P2}).} 
Note that in this approach neutrino states actually 
evolve only in time (see eq. (\ref{eq:P1})); the usual coordinate dependence 
of the oscillation probability (\ref{eq:P2}) is only obtained by invoking the 
additional ``time-to-space conversion'' assumption (\ref{eq:Lt}). Without 
this conversion, one would have come to a paradoxical conclusion that neutrino 
oscillations could be observed by just putting the neutrino detector 
immediately next to the source and waiting long enough.

Likewise, one could assume that all the mass-eigenstate neutrinos composing 
the initially produced flavour state $|\nu_a\rangle$ have the same energy. 
Using the fact that the spatial propagation of the $i$th mass eigenstate is 
described by the phase factor $e^{i {\bf p}_i{\bf x}}$ 
and that for a relativistic neutrino of mass $m_i$ and energy $E$  
\be
p_i=\sqrt{E^2-m_i^2}\simeq E-m_i^2/2E\,,
\label{eq:emr2}
\ee
one again comes to the same standard formula (\ref{eq:P2}) for the oscillation 
probability. Note that in this case the neutrino flavour evolution occurs in 
space and it is not necessary to invoke the ``time-to-space conversion'' 
relation (\ref{eq:Lt}) to obtain the standard oscillation formula.

The above two alternative derivations of the oscillation probability are 
very simple and transparent, and they allow one to arrive very quickly at 
the desired result. The trouble with them is that they are both wrong. 

In general, there is no reason whatsoever to assume that different mass 
eigenstates composing a flavour neutrino state emitted or absorbed in a 
weak-interaction process have either the same energy or the same momentum  
\footnote{The only exception we are aware of are neutrinos produced or 
detected in hypothetical M\"ossbauer-type experiments, since for them the 
``same energy'' assumption is indeed justified very well.}.
Indeed, the energies and momenta of particles emitted in any process are 
dictated by the kinematics of the process and by the experimental conditions.  
Direct analysis of, e.g., 2-body decays with simple kinematics, such as  
$\pi^\pm \to l^\pm + \nu_l(\bar{\nu}_l)$, where $l=e,\,\mu$, allows one to 
find the 4-momenta of the emitted particles and shows that neither energies 
nor momenta of the different neutrino mass eigenstates composing the flavour 
state $\nu_l$ are the same \cite{Wint,Giun1}. 
One might question this argument on the basis that it relies on the 
energy-momentum conservation and the assumption that the energies and momenta 
of the emitted mass eigenstates have well defined (sharp) values, whereas in 
reality these quantities have intrinsic quantum-mechanical uncertainties 
(see the discussion in Sec.~\ref{sec:disent}). However, the inexactness of the 
neutrino energies and momenta does not invalidate our argument that 
the ``same energy'' and ``same momentum'' assumptions are unjustified, and in 
fact only strengthens it. It should be also noted that the ``same energy'' 
assumption actually contradicts Lorentz invariance: even if it were satisfied 
in some reference frame (which is possible for two neutrino mass eigenstates, 
but not in the 3-species case), it would be violated in different Lorentz 
frames \cite{Giun1,Cohen1}. The same applies to the  ``same momentum'' 
assumption. 

One may naturally wonder why two completely different and wrong assumptions 
(``same energy''  and ``same momentum'') lead to exactly the same and 
correct result -- the standard oscillation formula. To understand that, 
it is necessary to consider the wave packet picture of neutrino oscillations. 
This will also allow us to analyze many other issues of the theory of neutrino 
oscillations. 

\section{\label{sec:why}Shape-independent wave packet approach to neutrino 
oscillations}

\subsection{\label{sec:wp} Neutrino wave packets and the oscillation 
probability}

In the discussion in the previous section we were actually considering 
neutrinos as plane waves or stationary states; strictly speaking, this 
description was inconsistent because such states are in fact non-propagating. 
Indeed, the probability of finding a particle described by a plane wave does 
not depend on the coordinate, while for stationary states this probability 
does not depend on time. In quantum theory propagating particles must be 
described by moving wave packets (see, e.g., \cite{PeskSchr}). We will 
therefore consider now the wave packet approach to neutrino oscillations. 
Many features of this approach have been discussed previously 
\cite{Nussinov:1976uw,Kayser:1981ye,Giunti:1991ca,Rich:1993wu,Dolgov:1997xr,
Giunti:1997wq,Cardall:1999ze,Dolgov:1999sp,Dolgov:2002wy,Giunti:2002xg,FarSm,
Visinelli:2008ds,nuwp2}. 
However, unlike in the majority of those papers, where Gaussian wave packets 
were considered, in our analysis we will not assume any specific shape 
of the wave packets. 

Let a flavour eigenstate $\nu_a$ be produced during a time interval $\Delta 
t_S$ centered at $t = 0$ in a source centered at ${\bf x}=0$. The wave packet 
describing the evolved neutrino state at a point with the coordinates 
$(t, {\bf x})$ is then 
\be
|\nu_a({\bf x}, t)\rangle\,=\,\sum_i U_{ai}^* \Psi_i({\bf x}, t)|\nu_i\rangle\,.
\label{eq:evolv}
\ee
We consider the evolution of the neutrino state (\ref{eq:evolv}) after the 
formation of the wave packet has been completed, that is at $t > \Delta t_S$. 
This state carries an information on the neutrino production process 
and corresponds to the mass-eigenstate components of the flavour neutrino 
state being on the mass shell (see Sec.~\ref{sec:size} for the discussion of 
this point). In what follows, we shall call it the production state. 
In eq.~(\ref{eq:evolv}), $\Psi_i({\bf x}, t)$ is the wave packet describing a 
free propagating neutrino of mass $m_i$:
\be
\Psi_i({\bf x},\, t)~=
\int\! \frac{d^3 p}{(2\pi)^{3/2}}\, f_i^S({\bf p}-{\bf p}_{i})
\,e^{i {\bf p} {\bf x} - i E_i(p) t}\,,
\label{eq:wp1}
\ee
where $f_i^S({\bf p}-{\bf p}_{i})$ is the momentum distribution function 
with ${\bf p}_{i}$ being the mean momentum, and $E_i(p)=\sqrt{p^2+m_i^2}$. 
The superscript $S$ at $f_i^S({\bf p}-{\bf p}_{i})$ indicates that the wave 
packet corresponds to the neutrino produced in the source. We will assume the 
function $f_i^S({\bf p}-{\bf p}_{i})$ to be sharply peaked at or very close 
to zero of its argument (${\bf p}={\bf p}_{i}$), with the width of the peak 
$\sigma_{pS}\ll p_{i}$. %
\footnote{For symmetric wave packets (i.e. when $f_i^S({\bf p}-{\bf p}_{i})$ 
is an even function of its argument), the position of the center of the peak 
coincides with the mean momentum ${\bf p}_{i}$. For asymmetric wave packets, 
it may be displaced from ${\bf p}_{i}$. }
No further properties of $f_i^S({\bf p}-{\bf p}_{i})$ need to be specified. 
Expanding $E_i(p)$ around the mean momentum, 
\be
E_i(p)=\left.E_i(p_{i})+\frac{\partial E_i(p)}{\partial p^j}
\right|_{{\bf p}_{i}}\!(p-p_{i})^j+\left.\frac{1}{2}\frac{\partial^2
E_i(p)}{\partial p^j \partial p^k}\right|_{{\bf p}_{i}}\!
(p-p_{i})^j\, (p-p_{i})^k +\dots\,,
\label{eq:emr3}
\ee
one can rewrite eq.~(\ref{eq:wp1}) as  
\be
\Psi_i({\bf x},\, t)\simeq e^{i{\bf p}_{i}{\bf x}-i E_i(p_i)t} \,
g_i^S({\bf x}-{\bf v}_{gi} t)\,,
\label{eq:wp2}
\ee
where
\be
g_i^S({\bf x}-{\bf v}_{gi} t)=\int\frac{d^3 p}{(2\pi)^{3/2}}\, 
f_i^S({\bf p})\,e^{i {\bf p}({\bf x}-{\bf v}_{gi} t)}\,
\label{eq:shape}
\ee
is the shape factor and  
\be
{\bf v}_{gi}~=~\left.\frac{\partial E_i}{\partial {\bf p}}\right|_{{\bf p}_i}~=
~\frac{{\bf p}}{E_i} \Big|_{{\bf p}_i} 
\label{eq:vg}
\ee
is the group velocity of the wave packet. Here we have retained only the 
first and the second terms in the expansion (\ref{eq:emr3}), since the higher 
order terms are of the second and higher order in the small neutrino mass and 
so can be safely neglected in all the situations of interest, with a possible 
exception of supernova neutrinos.  This approximation actually preserves the 
shape of the wave packets (and, in particular, neglects their spread). Indeed, 
the shape factor (\ref{eq:shape}) depends on time and coordinate only through 
the combination $({\bf x}-{\bf v}_{gi} t)$; this means that the wave packet 
propagates with the velocity ${\bf v}_{gi}$ without changing its shape. 

If the momentum dispersion corresponding to the momentum distribution 
function \mbox{$f_i^S({\bf p}-{\bf p}_i)$} is $\sigma_{pS}$, then, 
according to the Heisenberg uncertainty relation, the length of the wave 
packet in the coordinate space $\sigma_{xS}$ satisfies $\sigma_{xS}\gtrsim 
\sigma_{pS}^{-1}$; the shape factor function $g_i^S({\bf x}-{\bf v}_{gi} t)$ 
decreases rapidly when $|{\bf x}-{\bf v}_{gi} t|$ exceeds $\sigma_{xS}$.  
Eq.~(\ref{eq:wp2}) actually justifies and corrects the plane-wave approach: 
the wave function of a propagating mass-eigenstate neutrino is described by 
the plane wave corresponding to the mean momentum ${\bf p}_i$, multiplied 
by the shape function $g_i^S({\bf x}-{\bf v}_{gi} t)$ which makes sure 
that the wave is strongly suppressed outside a finite space-time region of 
width $\sigma_x$ around the point ${\bf x}={\bf v}_{gi} t$.

In the approximation where the wave packet spread is neglected, 
the evolved neutrino state is given by eq.~(\ref{eq:evolv}) with 
$\Psi_i({\bf x}, t)$ from eq.~(\ref{eq:wp2}). 
Note that the wave packets corresponding to different neutrino mass 
eigenstates $\nu_i$ are in general described by different momentum 
distribution functions $f_i^S({\bf p}-{\bf p}_{i})$ and therefore by 
different shape factors $g_i^S({\bf x}-{\bf v}_{gi} t)$. 

Let us now turn to the detected flavour-eigenstate neutrino $\nu_b$. We 
describe its state by a wave packet peaked at the coordinate $\bf{L}$ of the 
detecting particle: 
\be
|\nu_b({\bf x-L})\rangle =\sum_i U_{bi}^*\,\Psi_i^D({\bf x}-{\bf L})
|\nu_i\rangle\,.
\label{eq:PsiD}
\ee
This state has no time dependence because the detection process is essentially 
time independent on the time scale of the inverse energy resolution of 
the detector (see Sec.~\ref{sec:stat}).
The wave function $\Psi_i^D({\bf x}-{\bf L})$ can be written as 
\be
\Psi_i^D({\bf x}-{\bf L})~=
\int\! \frac{d^3 p}{(2\pi)^{3/2}}\, f_i^D({\bf p}-{\bf p}_{i}')
\,e^{i {\bf p} ({\bf x} -{\bf L})}\,,
\label{eq:wp1D}
\ee
where $f_i^D({\bf p}-{\bf p}_{i}')$ is the momentum distribution function 
of the wave packet characterizing the detection state, with ${\bf p}_i'$
being the mean momentum. Note that in general the average momenta of the 
production and detection states corresponding to the same neutrino mass 
eigenstate, ${\bf p}_i$ and ${\bf p}_i'$, are different.  
This is because the detection process may be sensitive to the interval 
of neutrino energies which does not exactly coincide with the energy spectrum 
of the emitted neutrino state 
(we shall discuss this point in more detail below). Just like for the 
production state, we will assume the function $f_i^D({\bf p}-{\bf p}_{i}')$ to 
be sharply peaked at or very close to zero of its argument (${\bf p}=
{\bf p}_{i}'$), with the width of the peak $\sigma_{pD}\ll p_{i}'$. 
By shifting the integration variable in eq.~(\ref{eq:wp1D}) one can  
rewrite $\Psi_i^D({\bf x}-{\bf L})$ as 
\be
\Psi_i^D({\bf x}-{\bf L}) = e^{i{\bf p}_{i}' ({\bf x}-{\bf L})} \,
g_i^D({\bf x}-{\bf L})\,,
\label{eq:wp2D}
\ee
with
\be
g_i^D({\bf x-L})=\int\frac{d^3 p}{(2\pi)^{3/2}}\, 
f_i^D({\bf p})\,e^{i {\bf p}({\bf x}-{\bf L})}\,.
\label{eq:shapeD}
\ee
The function $g_i^D({\bf x-L})$ is the shape factor of the wave packet 
corresponding to the detection of the $i$th mass eigenstate. It decreases 
quickly for $|{\bf x-L}|\gtrsim \sigma_{xD}$, where $\sigma_{xD}$ is the 
spatial length of the detection wave packet which is related to the momentum 
dispersion $\sigma_{pD}$ by the Heisenberg relation $\sigma_{xD}\gtrsim 
\sigma_{pD}^{-1}$.

The transition amplitude ${\cal A}_{ab}({\bf L}, t)$ describing neutrino 
oscillations is obtained by projecting the evolved state (\ref{eq:evolv}) onto 
(\ref{eq:PsiD}):
\be
{\cal A}_{ab}({\bf L}, t)=\int d^3 x \,\langle \nu_b({\bf x-L})|\nu_a({\bf x},t)
\rangle=\sum_i U^*_{ai} U_{bi}\,\int d^3 x\, \Psi_i^{D*}({\bf x-L}) 
\Psi_i({\bf x}, t)\,.
\label{eq:A}
\ee
Substituting here eqs.~(\ref{eq:wp2}) and (\ref{eq:wp2D}) yields 
\be
{\cal A}_{ab}({\bf L}, t)=
\sum_i U^*_{ai} U_{bi}\,G_i({\bf L-v}_{gi}t)\,
e^{-iE_i(p_i) t+i{\bf p}_i {\bf L}}\,,
\label{eq:A1}
\ee
where
\be
G_i({\bf L-v}_{gi}t)=\int d^3 x \,g_i^S({\bf x-v}_{gi}t)
g_i^{D*}({\bf x-L})\,e^{i {\bf (p}_i-{\bf p}_i') ({\bf x-L})}\,.
\label{eq:G}
\ee
That this integral indeed depends on ${\bf L}$ and ${\bf v}_{gi}t$ only 
through the combination $({\bf L-v}_{gi}t)$ can be easily shown by shifting 
the integration variable in (\ref{eq:G}). $G_i({\bf L-v}_{gi}t)$ is an 
effective shape factor whose width $\sigma_x$ depends on the widths of both 
the production and detection wave packets $\sigma_{xS}$ and $\sigma_{xD}$ 
and is of the order of the largest between them. 
Indeed, since the moduli of the shape factor functions $g_i^{S,D}$ quickly 
decrease when the arguments of these functions exceed the corresponding 
wave packet widths $\sigma_{xS}$ or $\sigma_{xD}$, from eq.~(\ref{eq:G}) it 
follows that $G_i({\bf L-v}_{gi}t)$ decreases when $|{\bf L-v}_{gi}t|$ 
becomes large compared to $max\{\sigma_{xS},\,\sigma_{xD}\}$. 
Actually, since $\sigma_x$ characterizes the overlap of the wave packets 
describing the production and detection states, it exceeds both $\sigma_{xS}$ 
and $\sigma_{xD}$. In particular, for Gaussian and Lorentzian (in the 
coordinate space) wave packets one has $\sigma_{x}=\sqrt{\sigma_{xS}^2+
\sigma_{xD}^2}$ and $\sigma_{x}=\sigma_{xS}+\sigma_{xD}$, respectively. 

The probability $P_{ab}({\bf L},\,t)\equiv P(\nu_a\to \nu_b;{\bf L},\,t)$ of 
finding a flavour eigenstate neutrino $\nu_b$ at the detector site at the 
time $t$ is given by the squared modulus of the amplitude ${\cal A}_{ab}
({\bf L}, t)$ defined in eq.~(\ref{eq:A}). Since in most experiments the 
neutrino emission and arrival times are not measured, the standard procedure 
in the wave packet approach to neutrino oscillations is then to integrate 
$P_{ab}({\bf L},\,t)$ over time. In doing so one has to introduce a 
normalization factor which is usually not calculated, 
\footnote{For an exception, see \cite{Cardall:1999ze}.} 
and in fact is determined by imposing ``by hand'' the requirement that the 
probabilities $P_{ab}(L)$ satisfy the unitarity condition. This is an 
{\it ad hoc} procedure which is not entirely consistent; the proper treatment 
would require to consider the temporal response function of the detector and 
would automatically lead to the correct normalization of the oscillation 
probabilities. 
It is shown in Sec.~\ref{sec:stat} that taking into account the temporal 
response function of the detector confirms 
the standard result,  and therefore we follow the same procedure here. The 
proper normalization of the oscillation probability is achieved by imposing 
the normalization condition 
\be
\int_{-\infty}^{\infty} dt\,|G_i({\bf L-v}_{gi}t)|^2  = 1\,.
\label{eq:norm}
\ee

For simplicity, from now on we neglect the transverse components of the 
neutrino momentum, i.e. the components orthogonal to the line connecting the 
centers of the neutrino source and detector; this is a very good approximation 
for neutrinos propagating macroscopic distances. The probability of finding a 
$\nu_b$ at the detector site provided that a $\nu_a$ was emitted by the source 
at the distance $L$ from the detector is then 
\be
P_{ab}(L) = \int_{-\infty}^{\infty} dt\,
|{\cal A}_{ab}(L, t)|^2 = \sum_{i,k} U^*_{ai} U_{bi} U_{ak} U_{bk}^*\,
I_{ik}(L)\,,
\label{eq:P3}
\ee
where
\be
I_{ik}(L)\equiv \int_{-\infty}^{\infty} dt\, G_i(L-v_{gi}t)\,
G_k^*(L-v_{gk}t)\,e^{-i\Delta \phi_{ik}(L,t)}\,. 
\label{eq:Iik1}
\ee
Here $G_i(L-v_{gi}t)$ is the effective shape factor corresponding to the $i$th 
neutrino mass eigenstate defined in (the 1-dimensional version of) 
eq.~(\ref{eq:G}). The quantity $\Delta\phi_{ik}(L,t)$ is the phase differences 
between the $i$th and $k$th mass eigenstates: 
\be
\Delta\phi_{ik}=(E_i-E_k)t -(p_i-p_k)L \;\equiv\; \Delta E_{ik} \,t ~-~ 
\Delta p_{ik} \, L\,,
\label{eq:phase1}
\ee
where
\be
E_i=\sqrt{p_i^2+m_i^2}\,.
\label{eq:emr}
\ee
Note that this phase difference is Lorentz invariant. 

To calculate the observable quantities -- the numbers of the neutrino 
detection events -- one has to integrate the oscillation probability (folded 
with the energy spectrum of the source, detection cross section and detector 
efficiency and energy resolution functions) over the neutrino spectrum and 
over the macroscopic volumes of the neutrino source and detector. 

\subsection{\label{sec:Iik} 
The standard oscillation formula and beyond}

According to eq.~(\ref{eq:P3}), the oscillation probability can be expressed 
through the elements of the leptonic mixing matrix and the integral 
$I_{ik}(L)$ defined in eq.~(\ref{eq:Iik1}). Let us now consider the 
properties of this integral. First, we derive a useful representation 
for the quantities $G_i(L-v_{gi}t)$. 
Expressing the shape factors $g_i^{S,D}(x)$ of the wave packets through the 
corresponding momentum distribution functions according to (the 1-dimensional 
versions of) eqs.~(\ref{eq:shape}) and (\ref{eq:shapeD}) and substituting the 
result into (\ref{eq:G}), we find 
\be
G_i(L-v_{gi}t)
=\int_{-\infty}^{\infty} dp\, f_i^S (p)
f_i^{D*}(p+\delta_i) e^{i p (L-v_{gi}t)}\,,
\label{eq:G1}
\ee
where
\be
\delta_i = p_i - p_i' 
\label{eq:deltai}
\ee
is the difference of the mean momenta of the production and detection states. 
If $\delta_i$ exceeds significantly the sum of the widths of the momentum 
distributions of the production and detection states $\sigma_{pS}+\sigma_{pD}$, 
the integral in (\ref{eq:G1}) is strongly suppressed due to the lack of 
overlap of the functions $f_i^S (p)$ and $f_i^{D*}(p+\delta_i)$ in the 
integrand. This is merely a manifestation of the approximate momentum 
conservation in the process: the mean momentum of the detected neutrino state  
should be approximately equal to that of the emitted neutrino, with possible 
deviation not exceeding the overall momentum uncertainty of the production and 
detection processes. Note that in the plane wave limit the momentum 
distribution functions $f_i^S (p-p_i)$ and $f_i^{D}(p-p_i')$ are reduced to 
the corresponding $\delta$-functions, and the momentum conservation becomes 
exact. In what follows we will always be assuming that $\delta_i$ satisfies 
$|\delta_i|\lesssim \sigma_{pS}+\sigma_{pD}$. 
 
Consider now the integral $I_{ik}(L)$. Substituting eqs.~(\ref{eq:G1}) 
and (\ref{eq:phase1}) into eq.~(\ref{eq:Iik1}) 
we find 
\bea
I_{ik}(L)=
\int_{-\infty}^{\infty} dt\, \int_{-\infty}^{\infty}
dp_1 \!\int_{-\infty}^{\infty} dp_2\,f_i^{S}(p_1) f_k^{D*}(p_1\!+\!\delta_i) 
f_i^{S*}(p_2) f_k^{D}(p_2\!+\!\delta_i) \nonumber \\
\times e^{i p_1 (L-v_{gi}t)-i p_2 (L-v_{gk}t)+i(\Delta p_{ik} L - 
\Delta E_{ik}t)}\,.
\label{eq:I1}
\eea
Performing first the integration over time and making use of the standard 
integral representation of Dirac's \mbox{$\delta$-function}, we obtain
\bea
&& \hspace*{-2.5cm}I_{ik}(L)=
e^{i (\Delta p_{ik}-\Delta E_{ik}/v_g)\,L}\;\frac{2\pi}{v_{gk}}
\int_{-\infty}^{\infty}\,dp\,f_i^{S}(p) f_i^{D*}(p+\delta_i) 
f_k^{S*}(r p+\Delta E_{ik}/v_{gk}) 
\nonumber \\
&& \hspace*{40mm} \times 
f_k^{D}(r 
p+\Delta E_{ik}/v_{gk}+\delta_i)\,e^{i p (1-r) L}\,,
\label{eq:I2}
\eea
where 
\be
v_g\equiv (v_{g_i}+v_{gk})/2\,,\qquad\qquad
r \equiv  \frac{v_{gi}}{v_{gk}}\simeq 1\,,~~~~~~~ 
\label{eq:r}
\ee
and we have neglected terms of order $(\Delta m_{ik}^2)^2$ in the phase factor 
in front of the integral. It can be readily shown that this factor 
actually contains the standard oscillation phase $(\Delta m_{ik}^2/2p) L$ 
(see eq.~(\ref{eq:exp1}) below), and thus we finally 
obtain 
\bea
&& \hspace*{-2.5cm}I_{ik}(L)=
e^{-i\frac{\Delta m_{ik}^2}{2p}\,L}
\;\frac{2\pi}{v_{gk}}
\int_{-\infty}^{\infty}\,dp\,f_i^{S}(p) f_i^{D*}(p+\delta_i) 
f_k^{S*}(r p+\Delta E_{ik}/v_{gk}) 
\nonumber \\
&& \hspace*{34mm} \times 
f_k^{D}(r 
p+\Delta E_{ik}/v_{gk}+\delta_i)\,e^{i p (1-r) L}\,. 
\label{eq:I3}
\eea
We shall now use this representation to discuss the properties of $I_{ik}(L)$. 
First, we notice that in the limit when the group velocities of the wave 
packets corresponding to different mass eigenstates are exactly equal to each 
other, $r=1$, the integral on the right hand side of eq.~(\ref{eq:I3}) does 
not depend on the distance $L$. Since the dominant contribution to this 
integral comes from the region $|p|\lesssim \sigma_P\equiv min\{\sigma_{pS},\,
\sigma_{pD}\}$ of the integration interval, the integral is practically 
independent of $L$ even for $r\ne 1$ provided that $|1-r|L\sigma_P\ll 1$, or 
\be
L\ll l_{\rm 
coh}=\sigma_X\frac{v_g}{\Delta v_g}\,,
\label{eq:coh1a}
\ee
where $\Delta v_g=|v_{gi}-v_{gk}|$ and $\sigma_X=1/\sigma_P$. This is merely  
the condition of the absence of the wave packet separation: the distance 
traveled by neutrinos should be smaller than the distance over which the wave 
packets corresponding to different mass eigenstates separate due to the 
difference of their group velocities and cease to overlap. If the condition 
opposite to that in eq.~(\ref{eq:coh1a}) is satisfied, the integral 
$I_{ik}(L)$ is strongly suppressed because of the fast oscillations of the 
factor $e^{i p (1-r) L}$ in the integrand. Thus, $I_{ik}(L)$ indicates how 
well the wave packets corresponding to the $i$th and $k$th neutrino mass 
eigenstates overlap with each other upon propagating the distance $L$ from 
the source.

In addition, $I_{ik}(L)$ is quenched if the split of the arguments of 
$f_i^{S,D}$ and $f_k^{S,D}$ in the integrand exceeds $\sigma_P$; therefore, a 
necessary condition for unsuppressed $I_{ik}(L)$ is 
\be
|\Delta E_{ik}|\sigma_X/v_g \ll 1\,.
\label{eq:loc1}
\ee 
As we shall show in Sec.~\ref{sec:coh1}, this condition is related to the 
coherence properties in the neutrino production and detection processes; 
therefore we shall call it the production/detection coherence condition 
or, for brevity, the interaction coherence condition.  

{}From the definition $\sigma_X=\sigma_P^{-1}\equiv [min\{\sigma_{pS},\,
\sigma_{pD}\}]^{-1}$ it follows that this quantity is of the order of 
$max\{\sigma_{xS},\,\sigma_{xD}\}$, i.e. of the same order of magnitude as the 
effective length $\sigma_x$ of the wave packets $G_i(L-v_{gi}t)$ introduced in 
the previous subsection. This, in particular, means that one can substitute 
$\sigma_x$ for $\sigma_X$ in the coherence conditions (\ref{eq:coh1a}) and 
(\ref{eq:loc1}). We shall be using this fact in what follows.

If the propagation and interaction coherence conditions 
(\ref{eq:coh1a}) and (\ref{eq:loc1}) are satisfied, one can set $r=1$ as well 
as neglect the quantity $\Delta E_{ik}$ in the integral in eq.~(\ref{eq:I3}). 
As shown in Appendix~A, the limit $r=1$ (i.e. $v_{gi}=v_{gk}$) also implies 
$f_i^{S,D}(p)=f_k^{S,D}(p)$. From the normalization condition (\ref{eq:norm})   
and eq.~(\ref{eq:G1}) it then follows directly that the integral in 
eq.~(\ref{eq:I3}) (with the factor $2\pi/v_{g}$ included) is simply equal to 
unity, so that $I_{ik}(L)=e^{-i(\Delta m_{ik}^2/2p) L}$. Substituting this 
into eq.~(\ref{eq:P3}) yields the standard oscillation probability.

\subsection{\label{sec:phase}Oscillation phase: answers to the questions}

Since neutrino oscillations occur due to the increasing phase difference 
between different neutrino mass eigenstates, one can learn a great deal about 
the oscillation phenomenon by studying the oscillation phase. We therefore 
concentrate on this phase now. This will allow us to answer some of the 
question raised in the Introduction and also to look at the results of the 
previous subsection from a slightly different viewpoint.   

Consider the phase difference $\Delta\phi_{ik}$ (\ref{eq:phase1}) that enters 
into the expression (\ref{eq:Iik1}) for $I_{ik}(L)$. To simplify the notation, 
we suppress the indices $i$ and $k$ where it cannot cause a confusion, so 
that $\Delta E\equiv \Delta E_{ik}$, $\Delta m^2\equiv \Delta m_{ik}^2$, etc. 
We shall consider the case $|\Delta E| \ll E$ which corresponds to relativistic 
or quasi-degenerate neutrinos. In this case one can expand the difference of 
the energies of two mass eigenstates in the differences of their momenta and 
masses. Retaining only the leading terms in this expansion, one gets 
\be
\Delta E=\frac{\partial E}{\partial p} \Delta p + 
\frac{\partial E}{\partial m^2} \Delta m^2 = v_g \,\Delta p  
+\frac{1}{2E}\,\Delta m^2\,,
\label{eq:exp1}
\ee
where $v_g$ is the average group velocity of the two mass eigenstates 
and $E$ is the average energy. Substituting this into eq.~(\ref{eq:phase1}) 
yields \cite{Sm1,Dolgov:1999sp,Dolgov:2002wy} 
\be
\Delta \phi=\frac{\Delta m^2}{2E}\, t ~-~ (L-v_g t) \Delta p \,.
\label{eq:phase2}
\ee
Note that our use of the mean group velocity and mean energy of the two mass 
eigenstates in eq.~(\ref{eq:phase2}) is fully legitimate. Indeed, going beyond 
this approximation would mean retaining terms of the second and higher order 
in $\Delta m^2$ in the expression for $\Delta\phi$. These terms are 
small compared to the leading ${\cal O}(\Delta m^2)$ terms; moreover, though 
their contribution to $\Delta\phi$ can become of order one at extremely 
long distances, the leading contribution to $\Delta \phi$ is then much greater 
than one, which means that neutrino oscillations are in the averaging regime 
and the precise value of the oscillation phase is irrelevant. 

Let us now consider the expression (\ref{eq:phase2}) for the phase difference 
$\Delta\phi$. If one adopts the same momentum assumption for the mean momenta 
of the wave packets representing the  different mass eigenstates, $\Delta 
p=0$, the second term on the right hand side 
disappears, which leads to the standard oscillation phase in the ``evolution 
in time'' picture. If, in addition, one assumes the ``time-to-space 
conversion'' relation (\ref{eq:Lt}), the standard formula for the 
$L$-dependent oscillation phase is obtained.

Alternatively, instead of expanding the energy difference of two mass 
eigenstates in the differences of their momenta and masses, one can expand 
the momentum difference of these states in the differences of their energies 
and masses: 
\be
\Delta p=\frac{\partial p}{\partial E} \Delta E + 
\frac{\partial p}{\partial m^2} \Delta m^2 = \frac{1}{v_g} \,\Delta E  
-\frac{1}{2p}\,\Delta m^2\,,
\label{eq:exp2}
\ee
where $p$ is the average momentum. 
Substituting this into eq.~(\ref{eq:phase1}) yields \cite{Sm2}   
\be
\Delta \phi = \frac{\Delta m^2}{2p} L ~-~ \frac{1}{v_g}(L-v_g t) \Delta E \,.
\label{eq:phase3}
\ee
Note that this relation could also be obtained directly from 
eq.~(\ref{eq:phase2}) by making use of eq.~(\ref{eq:exp1}). If one now adopts 
the same energy assumption for the mean energies of the wave packets, 
$\Delta E=0$, the second term on the right hand side vanishes, and one arrives 
at the standard oscillation phase. 

However, as we shall show now, eqs.~(\ref{eq:phase2}) and (\ref{eq:phase3})
actually lead to the standard oscillation phase even without the same energy 
or same momentum assumptions. For this purpose, let us generically write 
eqs.~(\ref{eq:phase2}) and (\ref{eq:phase3}) in the form 
\be
\Delta \phi = \Delta \phi_{st} + \Delta \phi'\,,
\label{eq:phi1}
\ee
where $\Delta \phi_{st}$ is the standard oscillation phase either in 
``evolution in time'' or in ``evolution in space'' approach, and $\Delta 
\phi'$ is the additional term (the second term in eq.~(\ref{eq:phase2}) or 
(\ref{eq:phase3})). The first thing to notice is that $\Delta \phi'$ vanishes 
not only when $\Delta p=0$ (in eq.~(\ref{eq:phase2})) or $\Delta E=0$
(in eq.~\ref{eq:phase3})), but also at the center of the wave packet, where 
$L=v_{g} t$. Away from the center, the quantity $L-v_g t$ does not vanish, but 
it never exceeds substantially the length of the wave packet $\sigma_x$, since 
otherwise the shape factors would strongly suppress the neutrino wave 
function; thus, $|L-v_g t| \lesssim \sigma_x$. The physical meaning of the two 
terms in eq.~(\ref{eq:phi1}) is then clear: $\Delta \phi_{st}$ is the 
oscillation phase acquired by the neutrino state over the distance $L$ 
for a pointlike neutrino, whereas $\Delta \phi'$ takes into account that the 
wave packet actually has a finite size and is the additional phase variation 
along the wave packet. 
{}From eq.~(\ref{eq:phase3}) it is clear that the term $\Delta \phi'$ 
can be neglected when the spatial length of the wave packet $\sigma_x$ is 
small compared $v_g/\Delta E$. This coincides with the interaction coherence 
condition (\ref{eq:loc1}) considered in Sec.~\ref{sec:Iik}. 
We will discuss this condition and the coherence properties of the neutrino 
production and detection processes in more detail in Sec.~\ref{sec:coh1}.

Let us now show explicitly that under very general assumptions the neutrino 
oscillation probability takes its standard form (\ref{eq:P2}). We start with 
the expression for 
$\Delta\phi$ in eq.~(\ref{eq:phase3}). Substituting it into (\ref{eq:Iik1}), 
we find 
\bea
I_{ik}(L) = e^{-i\frac{\Delta m_{ik}^2}{2p}\,L}\int_{-\infty}^{\infty} dt\, 
G_i(L-v_{gi}t)\,G_k^*(L-v_{gk}t) \,e^{i\frac{1}{v_g}\Delta 
E_{ik}(L-v_g t)} \,. 
\label{eq:Iik2}
\eea
Let us first neglect the difference between the group velocities of the 
wave packets describing different mass eigenstates, i.e. take $v_{gi}=v_{gk}
=v_g$, which also implies  $G_k=G_i$ (see Appendix A). In this approximation,  
which neglects the decoherence effects due to the wave packet separation, 
the integral in (\ref{eq:Iik2}) does not depend on $L$; this can be readily 
shown by changing the integration variable according to $t\to (L-v_{g}t)$. 
The same conclusion has already been reached in Sec.~\ref{sec:Iik} using the 
momentum-integral representation of $I_{ik}(L)$. The integral in 
eq.~(\ref{eq:Iik2}) is then just the Fourier transform of the 
squared modulus of the shape factor: 
\be
\frac{1}{v_g}\int_{-\infty}^{\infty} dx'\, 
|G_i(x')|^2 \,e^{i\frac{1}{v_g}\Delta E_{ik}x'} \,. 
\label{eq:Iik3}
\ee
It is essentially a coherence factor, which takes into account the effects of 
suppression of the oscillations in the case when the energy and/or momentum 
uncertainties at neutrino production or detection are small enough to allow 
the determination of the neutrino mass (see Sec.~\ref{sec:coh1} for a more 
detailed discussion). If the interaction coherence condition 
(\ref{eq:loc1}) is fulfilled, one can replace the oscillating phase factor 
in the integrand of eq.~(\ref{eq:Iik3}) by unity; the resulting integral is 
then simply equal to 1, as follows from the normalization condition 
(\ref{eq:norm}). Eq.~(\ref{eq:P3}) then immediately leads to the standard 
oscillation probability (\ref{eq:P2}). If, on the contrary, 
the condition opposite to the interaction coherence condition 
(\ref{eq:loc1}) is satisfied, the integral (\ref{eq:Iik3}) is strongly 
suppressed due to the fast oscillations of the factor $\exp(i\Delta E_{ik}x'
/v_g)$ in the integrand, leading to the suppression of the oscillations. In 
the borderline case $\Delta E_{ik}\sigma_x/v_g\sim 1$, a partial 
decoherence occurs. 

If one now allows for $v_{gi}\ne v_{gk}$, then, as direct inspection of the 
arguments of the functions $G_{i,k}$ in the integrand in eq.~(\ref{eq:Iik2}) 
shows, the dependence of the integral in this equation on $L$ is still 
negligible if $L\Delta v_g /v_g\ll \sigma_x$, i.e. if the condition 
(\ref{eq:coh1a}) is satisfied. As has already been pointed out, this is the 
condition of the absence of decoherence due to the wave packet separation: 
the distance traveled by neutrinos should be smaller than the distance over 
which the wave packets corresponding to different mass eigenstates separate, 
due to the difference of their group velocities, to such an extent that 
their effects can no longer interfere in the detector.%
\footnote{We reiterate that $\sigma_x$ is an {\sl effective} spatial length 
of the wave packet, which depends on the lengths of both the production and 
detection wave packets $\sigma_{xS}$ and $\sigma_{xD}$. The condition 
(\ref{eq:coh1a}) therefore automatically takes into account possible 
restoration of coherence at detection, as discussed in Sec.~\ref{sec:sep}.} 
If the condition opposite to that in eq.~(\ref{eq:coh1a}) is satisfied, the 
integral in eq.~(\ref{eq:Iik2}) is strongly suppressed because of the lack of 
overlap between the factors $G_i(L-v_{gi}t)$ and $G_k^*(L-v_{gk}t)$ in the 
integrand (if $I_{ik}(L)$ is written as a momentum-space integral, the 
suppression is due to the fast oscillations of the integrand, see 
eq.~(\ref{eq:I3})). 

{}From the above consideration it follows that the factor $I_{ik}(L)$ in the 
expression for the oscillation probability (\ref{eq:P3}) yields the standard 
oscillation phase factor $\exp{(-i\frac{\Delta m_{ik}^2}{2p}\,L)}$ multiplied 
by the integral which accounts for possible suppression of the oscillating 
terms due to decoherence caused by the wave packet separation and/or lack of 
coherence at neutrino production or detection. Note that both decoherence 
mechanisms usually lead to exponential suppression of the interference terms 
in the oscillation probabilities since they come from the infinite-limits 
integrals of fast oscillating functions. The exact form of these suppression 
factors depends on the shape of the wave packets, i.e. is model dependent; in 
particular, for Gaussian and Lorentzian wave packets, these factors are 
$\,\sim\exp[-(L/l_{\rm coh})^2] \exp[-(\sigma_x \Delta E)^2/2v_g^2]$ and 
$\exp(-L/l_{\rm coh})\exp(-\sigma_x \Delta E/2 v_g)$, respectively.

Thus we conclude that {\sl the standard oscillation probability is obtained 
if neutrinos are relativistic or quasi-degenerate in mass and the decoherence 
effects due to the wave packet separation or lack of coherence at neutrino 
production or detection are negligible. No unjustified ``same energy'' or 
``same momentum'' assumptions are necessary to arrive at this result.}

One may wonder why these assumptions are actually so popular in the literature 
and even made their way to some textbooks, though in general there is no good 
reason to believe that the different mass eigenstates have either same 
momentum or same energy. One possible reason could be the simplicity of the 
derivation of the formula for the oscillation probability under these 
assumptions. However, we believe that simplicity is no justification for using 
a wrong argument to arrive at the correct result.

\section{\label{sec:another}Another standpoint}

In this section we outline a more general approach to the calculation  
of the oscillation probability, which gives an additional insight into the 
issues discussed in the present paper. Here we just illustrate some points 
relevant to our discussion rather than presenting a complete formalism, which 
is beyond the scope of our paper.

The wave functions of the production and detection states $\Psi_i^S (x,t)$ and 
$\Psi_i^D(x-L)$ can be represented as the integrals over momenta according 
to (the 1-dimensional versions of) eqs.~(\ref{eq:wp1}) and ~(\ref{eq:wp1D}). 
Inserting these expressions into eq.~(\ref{eq:A}) and performing the integral 
over the coordinate, we obtain   
\be
{\cal A}_{ab}(L, t) =  \sum_i  U_{a i}^* U_{bi}\int
dq_i \, h_i(q_{i}) e^{i q_i L - i E_i(q_i) t}\,,
\label{eq:ampl-p}
\ee 
where 
\be
h_i(q) \equiv   f_i^S (q-p_i)  f_i^{D*}(q-p_i')\,.
\label{eq:h1}
\ee
Here the momenta $q_i$ are the integration variables, whereas $p_i$ and 
$p_i'$ are, as before, the mean momenta of the corresponding wave packets. 

The amplitude (\ref{eq:ampl-p}) is the sum of plane waves corresponding to 
different momenta and different masses. The integration over momenta can be 
formally substituted by a summation to make this point clearer. The oscillation 
probability is then  
\be
P_{ab}(L, t) \equiv |{\cal A}_{ab}(L, t)|^2=
\left|\sum_i \sum_{q_i} U_{a i}^* U_{bi} 
h_i(q_{i}) e^{i q_i L - i E_i(q_i) t}\right|^2 . 
\label{eq:prob}
\ee
The waves with all momenta and masses should be summed up; the resulting 
expression for the oscillation probability includes the interference of 
these waves.

The standard approach to the calculation of the oscillation amplitude  
(\ref{eq:ampl-p}) (or probability (\ref{eq:prob})) is to sum up first the 
waves with different momenta but the same mass, and then sum over the mass 
eigenstates. In this way first the wave packets corresponding to different 
mass eigenstates are formed, and then the interference of these wave packets 
is considered. 
Since
\be 
\int
 dq_i\, h_{i}(q_{i}) e^{i q_i L - i E_i(q_i) t}
=  G_i (L - v_{gi}t) e^{i p_i L - i E_i(p_i) t}\,, 
\ee
eq.~(\ref{eq:ampl-p}) then directly reproduces the results of 
Sec.~\ref{sec:why}. 

Another possibility is to sum up first the plane waves with different masses 
but equal (or related) momenta and then perform the integration over the 
momenta. In particular, one can select the waves with equal energies. Clearly, 
the final result should not depend on the order of summation if no 
approximations are made, and should be almost independent of this order if the 
approximations are well justified. However, different summation conventions 
allow different physical interpretations of the result. In what follows we 
will perform computations using the ``equal momenta'' and ``equal energy'' 
summation rules and identify the conditions under which they lead to the 
standard result for the oscillation probability. 

Let us first consider the ``equal momenta'' summation. Setting $q_i=p$ for 
all $i$, one can write the amplitude (\ref{eq:ampl-p}) as  
\be
{\cal A}_{ab}(L, t) = \int dp 
\sum_i U_{a i}^* U_{bi}\, h_{i}(p)  e^{i p L - i E_i(p) t}.  
\label{eq:ampl-pp}
\ee 
{}From eq.~(\ref{eq:h}) and the definition (\ref{eq:h1}) of $h_i(p)$ it 
follows that $h_{i}(p) = h(p, E_i(p))$. We can use this to  
expand $h_i$ in a power series in $\Delta m_{i3}^2$: 
\be
h_{i}(p)= h(p, E_i) = h(p, E_3) + \left. E_i\frac{\partial h_i}{\partial E_i}
\right|_{m_i = m_3}\frac{\Delta m^2_{i3}}{2E_3^2} + ...
\ee
Inserting this expression into (\ref{eq:ampl-pp}), we obtain 
\be
{\cal A}_{ab}(L, t) =  \int dp\,  h (p, E_3(p)) e^{i p L - i E_3(p) t}
\sum_i U_{a i}^* U_{bi}  e^{- i [E_i(p) - E_3(p)]t} + {\cal A}^{\Delta}_{ab}
(L, t)\,, 
\label{eq:ampl-pp1}
\ee
where 
\be
{\cal A}_{ab}^\Delta (L, t)
 \equiv \int dp\, e^{i p L - i E_3(p) t} \sum_i U_{a i}^* U_{bi} e^{- 
i [E_i(p) - E_3(p)]t} \left.\frac{E_i\partial h_i} {\partial 
E_i}\right|_{m_i=m_3} \,\frac{\Delta m^2_{i3}}{2E_3^2(p)}. 
\label{eq:ampl-pp2} 
\ee 
It is easy to see that the term ${\cal A}_{ab}^\Delta (L, t)$ is typically 
very small: 
\be
{\cal A}_{ab}^\Delta (L, t)\sim \frac{\Delta m^2_{i3}}{E\sigma_E}\,
{\cal A}_1\,, 
\label{eq:est1}
\ee
where ${\cal A}_1$ is the first term on the right hand side of 
eq.~(\ref{eq:ampl-pp1}). Indeed, if the width of the effective momentum 
distribution function $h(p)$ is $\sigma_p$, one has $\partial h/\partial 
p\sim h/\sigma_p$, so that 
\be 
\frac{E\partial h}{\partial E}\,=\,\frac{E}{v_g}\frac{\partial h}
{\partial p} \,\sim\, \frac{Eh}{v_g\sigma_p}\,=\, 
\frac{Eh}{\sigma_E}\,,
\label{eq:est2} 
\ee 
which immediately leads to (\ref{eq:est1}). Thus, if
\be 
\sigma_E\gg \frac{\Delta m^2}{2E}\,,
\label{eq:loc3}
\ee
${\cal A}_{ab}^\Delta (L, t)$ can safely be neglected. 

Consider now the first term in eq.~(\ref{eq:ampl-pp1}). 
For a fixed momentum the phase difference is $\Delta\phi_{i3} = (E_i - E_3)t 
\approx \Delta m^2_{i3}t/ 2 E_3$.%
\footnote{This approximation breaks down at very small momenta. Note, 
however, that the small-$p$ contribution to the integral in 
(\ref{eq:ampl-pp1}) is strongly suppressed because of the effective 
momentum distribution function $h(p,E_3)$, which is strongly peaked at 
a relativistic momentum $p=p_3$. This justifies using the 
approximation for $\Delta \phi_{i3}$ in (\ref{eq:ampl-pp1}).} If the 
width of the effective momentum distribution function $h (p, E_3)$ is 
small enough, so that the change of the phase within the wave packet 
is small, we can pull the oscillatory factor out of the integral at 
some effective momentum (corresponding to an energy $E$): 
\bea
{\cal A}_{ab}(L, t) 
\,\simeq \left[\sum_i U_{a i}^* U_{bi} 
e^{- i \frac{\Delta m^2_{i3}}{2 E} t}\right]\int dp\, h (p, E_3(p)) 
e^{i p L - i E_3(p) t} 
\nonumber \\
= \left[\sum_i U_{a i}^* U_{bi} e^{- i \frac{\Delta m^2_{i3}}{2 E} 
t}\right]
G_3(L-v_{g3}t)e^{i p_3 L - i E_3(p_3) t} \,. ~\,
\label{eq:ampl} 
\eea
Here the factor in the square brackets gives the standard oscillation 
amplitude in the ``evolution in time'' approach. Integrating the squared 
modulus of the amplitude (\ref{eq:ampl}) over time and using once again 
eq.~(\ref{eq:loc3}), one arrives at the standard expression for the 
oscillation probability. 

Thus, we obtain the standard oscillation formula by first summing up the 
waves with equal momenta and different masses and then integrating over the 
momenta provided that the following two conditions are satisfied: 
\begin{itemize}
\item[(i)] The variation of the oscillation phase within the wave packet due 
to the energy spread is small: $\sigma_E \ll (2\pi E^2/\Delta m^2_{ik}) 
L^{-1}$; this condition allows one to pull the oscillatory factor out of the 
integral over the momenta, as discussed above. Note that it is actually 
equivalent to the condition of no wave packet separation, 
eq.~(\ref{eq:coh1a}) (recall that $\Delta v_g\simeq \Delta m^2/2E^2$ and  
$\sigma_x\simeq v_g/\sigma_E$).  

\item[(ii)] 
The momentum distribution functions $h_i(p)$ are not too narrow: 
$\sigma_E\gg \Delta m^2/2E$. This condition, in particular, allows one 
to neglect the contribution ${\cal A}_{ab}^\Delta (L, t)$ in 
eq.~(\ref{eq:ampl-pp1}). 
It actually ensures that the neutrino wave packet length $\sigma_x\simeq 
v_g/\sigma_E$ is small compared to the neutrino oscillation length 
and is related to the interaction coherence condition, as we shall show in 
Sec.~\ref{sec:coh1}.

\end{itemize}
These conditions for obtaining the standard oscillation formula essentially 
coincide with the conditions found in a different framework in 
Sec.~\ref{sec:phase}. 

To describe the possible decoherence effects due to the separation of 
the wave packets and lack of coherence at neutrino production or 
detection explicitly, one should lift the  conditions (i) and (ii) and 
consider the corresponding corrections to the oscillation amplitude. 
This is discussed in detail in Sec. \ref{sec:QM}. 

Similarly, we can consider the summation of waves with equal energies and 
different masses, with the subsequent integration over energies (or momenta). 
Requiring $E_i(q_i) = E_3(p)$ yields $q_i = \pm \sqrt{p^2+\Delta m^2_{3i}}$ 
$(i=1, 2)$, $q_3=p$. Taking into account that $d q_i = \pm dp/\sqrt{1 + \Delta 
m^2_{3i}/p^2}$, we obtain 
\be
{\cal A}_{ab}(L, t) =  \int dp\, 
\sum_i U_{a i}^* U_{bi} \frac{h_i(q_i(p), E_3(p))}{\sqrt{1 + \Delta 
m^2_{3i}/p^2}} e^{i q_i(p) L - i E_3(p) t}.  
\label{eq:ampl-e}
\ee 
Here it is assumed that $\nu_3$ is the heaviest mass eigenstate, so that 
$\Delta m_{3i}^2\ge 0$ and no singularities appear in the integrand. 
Note that our change of the integration variables $q_i\to p$  
excludes, for $i=1, 2$, the small regions of momenta $q_i$ around zero; this, 
however, introduces only a tiny error, because the main contributions to the 
integral come from the regions around the points $q_i=p_i$, where the 
functions $h_i(q_i)$ are strongly peaked. 
 
Expanding $h_i(q_i(p),E_3(p))$ as 
\be
\frac{h_{i}(q_i, E_3(p))}{\sqrt{1 + \Delta m^2_{3i}/p^2}} \simeq  
h(p, E_3(p))  + \left(h(p, E_3(p)) - p\frac{\partial h(p,E_3(p))}{\partial p}
\right) \frac{\Delta m^2_{3i}}{2p^2}
\ee
and inserting this expression into (\ref{eq:ampl-e}), we obtain  
\be
{\cal A}_{ab}(L, t)  =  \int dp\,  h (p, E_3(p))  e^{i p L - i 
E_3(p) t}\sum_i U_{a i}^* U_{bi}  e^{i [q_i(p) - p]L}  +  
\bar{A}_{ab}^{\Delta}, 
\ee
where 
\be
\bar{{\cal A}}_{ab}^\Delta (L, t) 
 \equiv \int dp\,  
e^{i p L - i E_3(p) t}\sum_i U_{a i}^* U_{bi} e^{i [q_i(p) - p]L} 
\left(h-\frac{p \partial h}{\partial p}\right)
\frac{\Delta m^2_{3i}}{2p^2}\,. 
\label{eq:ampl-pp3}
\ee 
Just as in the previous case, one can show that the amplitude 
$\bar{{\cal A}}_{ab}^\Delta (L, t)$ can be neglected if the condition 
(\ref{eq:loc3}) is satisfied. Assuming this to be the case and that the 
variation of the oscillation phase within the wave packet due to the momentum 
spread is small, and taking into account that $(q_i - p)L=(q_i - q_3)L\simeq 
\Delta m_{3i}^2L/2p$, one finds 
\be
{\cal A}_{ab}(L, t)\, \simeq 
\left[\sum_i U_{a i}^* U_{bi} e^{i \frac{\Delta m^2_{3i}}{2 p} 
L}\right]
G_3(L-v_{g3}t)e^{i p_3 L - i E_3(p_3) t} \,,
\label{eq:ampl-pp4}
\ee
where $p$ is the average neutrino momentum. 
This is the standard oscillation amplitude multiplied by the effective shape 
factor of the wave packet of $\nu_3$ (note that in our current approximation 
we actually neglect the difference between the wave packets of different mass 
eigenstates). Integrating the squared modulus of the amplitude in 
eq.~(\ref{eq:ampl-pp4}) over time and using the normalization condition 
(\ref{eq:norm}), we again arrive at the standard expression for the 
oscillation probability, just as in the previous case when we first summed 
the terms with equal momenta and different masses and then integrated over the 
momenta. In deriving this result we once again used the conditions (i) and (ii) 
discussed above. 

Our discussion of the new summation rules for calculating the oscillation 
probability presented here leads to an alternative explanation of why the 
``same energy'' and ``same momentum'' assumptions eventually lead to the 
correct physical observables, as discussed in Sec.~\ref{sec:stat}.

\section{\label{sec:QM}Quantum-mechanical uncertainty relations and 
neutrino oscillations} 

Neutrino oscillations, being a quantum-mechanical interference phenomenon, 
owe their very existence to quantum-mechanical uncertainty relations.
The coordinate-momentum and time-energy uncertainty relations are implicated 
in the oscillations phenomenon in a number of ways. First, it is the energy 
and momentum uncertainties of the emitted neutrino state that allow it  
to be a coherent superposition of the states of 
well-defined and different mass. The same applies to the detection process -- 
for neutrino detection to be coherent, the energy and momentum uncertainties 
inherent in the detection process should be large enough to prevent a 
determination of the absorbed neutrino's mass in this process. 
The uncertainty relations also determine the size of the neutrino wave 
packets and therefore are crucial to the issue of the loss of coherence 
due to the wave packet separation. In addition, these relations are 
important for understanding how the produced and detected neutrino states 
are disentangled from the accompanying particles. Let us now discuss these 
issues in more detail. We start with 

\subsection{\label{sec:disent}Uncertainty relations and disentanglement 
of neutrino states }

In the majority of analyses of elementary particle processes it is assumed 
that the energies and momenta of all the involved particles have well defined 
(sharp) values and obey the exact conservation laws. However, for this 
description to be exact, the considered processes (and the particles involved) 
should be completely delocalized in space and in time, whereas in reality 
these processes occur in finite and relatively small spatial volumes and 
during finite time intervals. For this reason, the energy and momenta of all 
the participating particles have intrinsic quantum mechanical uncertainties, 
and the particles should be described by wave packets rather than states 
of definite momentum -- plane waves (see, e.g., \cite{PeskSchr}). The 
conservation of energy and momentum for these particles is also fulfilled up 
to these small uncertainties. 

This does not, of course, mean that the energy-momentum conservation, which is 
a fundamental law of nature, is violated: it is satisfied exactly when one 
applies it to all particles in the system, including those whose interactions  
with the particles directly involved in the process localize the latter in a 
given space-time region. Schematically speaking, if we consider the process as 
occurring in a box, the interactions with the walls of the box and the 
contributions of these walls to the energy-momentum balance have to be taken 
into account. In practice, this is never done; however, the resulting 
inaccuracy of the energy and momentum conservation as well as the intrinsic 
quantum-mechanical uncertainties of the energies and momenta of the involved 
particles are usually completely negligible compared to their energies and 
momenta themselves, and therefore can be safely ignored in most processes.

This is, however, not justified when neutrino oscillations are considered, 
since the neutrino energy and momentum uncertainties, as tiny as they are, are 
crucially important for the oscillation phenomenon. In this respect, we 
believe that the attempts to use the exact energy-momentum conservation in the 
analyses of neutrino oscillations are inconsistent. In some analyses 
the exact energy-momentum conservation is 
assumed for the  neutrino production and detection processes in order to 
describe neutrinos as being entangled with accompanying particles.  The  
subsequent disentanglement, which is necessary for neutrino oscillations to 
occur, is assumed to be due to the interaction of these accompanying particles 
(such as e.g. electrons or muons produced in decays of charged pions) with 
medium. This localizes those particles and creates the necessary energy and 
momentum uncertainties for the neutrino state. The described approach misses 
the fact that the parent particles are already localized in the neutrino 
production and detection processes, and so no additional disentanglement 
through the interaction of the accompanying particles with medium is 
necessary. Indeed, it is clear that neutrinos produced, for example, in 
$\pi^\pm$ decays oscillate even if the accompanying charged leptons do not 
interact with medium, i.e. are not ``measured''. The measurement of the 
flavour of these charged leptons that discriminates between $e^\pm$ and 
$\mu^\pm$ and makes neutrino oscillations possible is actually provided by 
the decoherence of the charged leptons due to their very large mass difference 
\cite{Akh2}. 

\subsection{\label{sec:size}What determines the size of the wave 
packet?}

According to the quantum-mechanical uncertainty relations, the energy and 
momentum uncertainties of a neutrino produced in some process are determined 
by, correspondingly, the time scale of the process and spatial localization 
of the emitter. These two quantities are in general independent; on the other 
hand, for a free on-shell particle of definite mass the dispersion relation 
$E^2=p^2+m^2$ immediately leads to 
\be
E \sigma_E = p \sigma_p\,.
\label{eq:rel1}
\ee
Since this relation is satisfied for each mass-eigenstate component of 
the emitted flavour state, it must also be satisfied for the state as 
a whole (provided that the energies and momenta of different components 
as well as their uncertainties are nearly the same, which is the case 
for relativistic or quasi-degenerate neutrinos). Thus, we have an apparently 
paradoxical situation: on the one hand, $\sigma_E$ and $\sigma_p$ should be 
essentially independent, while on the other hand they must satisfy 
eq.~(\ref{eq:rel1}). 

The resolution of this paradox comes from the observation that at the time of 
their production neutrinos are actually not on the mass shell and therefore do 
not satisfy the standard dispersion relation. Therefore their energy and 
momentum uncertainties need not satisfy (\ref{eq:rel1}). However, as soon as 
neutrinos move away from their production point and propagate distances $x$ 
such that $p x\gg 1$, they actually go on the mass shell, and their energy and 
momentum uncertainties start obeying eq.~(\ref{eq:rel1}). This happens because 
the bigger of the two uncertainties shrinks towards the smaller one, so that 
eq.~(\ref{eq:rel1}) gets fulfilled. Indeed, when the neutrinos go on the mass 
shell, the standard relativistic dispersion relation connecting the energies 
and momenta of their mass eigenstate components allows to determine the less 
certain of these two quantities through the more certain one, thus reducing the 
uncertainty of the former. As a result, the two uncertainties get related by 
eq.~(\ref{eq:rel1}), with the one that was smaller at production retaining its 
value also on the mass shell. Note that for neutrino energies in the MeV 
range neutrinos go on the mass shell as soon as they propagate distances 
$x\gtrsim 10^{-10}$~cm from their birthplace.

Which of the two uncertainties, $\sigma_p$ or $\sigma_E$, is actually the 
smaller one at production? Quite generally, this happens to be the energy 
uncertainty 
$\sigma_E$. Indeed, consider, e.g., an unstable particle, the decay of which 
produces a neutrino. In  reality, such particles are always localized in 
space, so one can consider them to be confined in a box of a linear size 
$L_S$. The localizing ``box'' is actually created by the interactions of the 
particle in question with the surrounding particles. Assume first that the 
average time interval $T_S$ between two subsequent collisions of the decaying 
particle with the walls of the box (more precisely, the interval between its 
collisions with the surrounding particles) is shorter than its lifetime 
$\tau=\Gamma^{-1}$. Then the energy width of the state produced in the decay 
is given by the so-called collisional broadening, and is actually $\simeq 
T_S^{-1}$. This width directly gives the neutrino energy uncertainty, i.e. 
$\sigma_E\simeq T_S^{-1}$. On the other hand, the neutrino momentum uncertainty 
is $\sigma_p\simeq L_S^{-1}$. Since $T_S$ is related to $L_S$ through the 
velocity of the parent particle $v$ as $T_S\simeq  L_S/v$, one finds
\be
\sigma_E < \sigma_p\,,
\label{eq:ineq1}
\ee
which is actually a consequence of $v<1$. 

Consider now the situation when the lifetime of the parent particle is shorter 
than the interval between two nearest collisions with the walls of the box. In 
this case the decaying particle can be considered quasi-free, and the energy 
uncertainty of the produced neutrino is given by the decay width of the parent 
particle: $\sigma_E\simeq \Gamma$.%
\footnote{This only holds for slow parent particles. In the case of 
relativistic decaying particles, $\sigma_E$ depends on the angle between the 
momenta of the parent particle and of neutrino, see \cite{FarSm} and 
Sec.~\ref{sec:Lorentz} below. Still, the condition (\ref{eq:ineq1}) holds in 
that case as well.}
The momentum uncertainty of neutrino is 
then the reciprocal of its coordinate uncertainty $\sigma_x$, which in turn is 
just the distance traveled by neutrino during the decay process: $\sigma_x
\simeq (p/E)\tau=(p/E)\Gamma^{-1}\simeq (p/E)\sigma_E^{-1}$. Thus we find 
$p \sigma_p \simeq E \sigma_E$, i.e. eq.~(\ref{eq:rel1}) is approximately 
satisfied in this case. Once again the condition (\ref{eq:ineq1}) is fulfilled. 
It can be shown that this inequality is also satisfied when neutrinos are 
produced in collisions rather than in decays of unstable particles 
\cite{Beuthe2}. Similar arguments apply to the neutrino detection process.  
Actually, by $\sigma_E$ and $\sigma_p$ in the above discussion (as 
well as in the discussion in Sec.~\ref{sec:coh1}) one should understand 
the effective energy and momentum uncertainties, which depend on the 
corresponding uncertainties both at neutrino production and detection 
and are dominated, both for the neutrino energy and momentum, by the 
smallest between the production and detection uncertainties.

Thus, we conclude that the energy uncertainties at neutrino production and 
detection are always smaller than the corresponding momentum uncertainties. 
This has important implications for the effective neutrino wave packets 
$G_i(L-v_{gi}t)$. As discussed in Sec.~\ref{sec:wp}, they are characterized 
by the effective momentum uncertainty $\sigma_p^{\rm eff}\sim \sigma_P\equiv 
min\{\sigma_{pS},\,\sigma_{pD}\}$, and similarly for the effective energy 
uncertainty. Since the wave packets describe propagating on-shell particles, 
from the above discussion it follows that the effective energy uncertainty 
$\sigma_E^{\rm eff}$ characterizing $G_i(L-v_{gi}t)$ coincides with the 
energy uncertainty at production/detection discussed above: 
$\sigma_E^{\rm eff}=\sigma_E$. At the same time, the effective momentum 
uncertainty of the wave packet $\sigma_p^{\rm eff}$ has in general nothing to 
do with the momentum uncertainties inherent in the neutrino production and 
detection processes; rather, it is related to the energy uncertainty 
$\sigma_E$ through the on-shell condition (\ref{eq:rel1}): $\sigma_p^{\rm eff}
=\sigma_E/v_g$. Because the spatial length of the wave packets is $\sigma_x\sim 
1/\sigma_p^{\rm eff}$, this in turn means that  
\be
\sigma_x \sim v_g/\sigma_E\,.
\label{eq:sigmax3}
\ee
Thus, we conclude that the spatial length $\sigma_x$ of the wave packets 
describing the propagating neutrino states is always determined by the energy 
uncertainty at neutrino production/detection as the smaller one between 
$\sigma_p$ and $\sigma_E$. This is in accord with the known fact that for 
stationary neutrino sources (for which $\sigma_E=0$) the neutrino coherence 
length is infinite \cite{Kiers,Grimus:1996av}. On the other hand, as 
will be shown in the next subsection, the localization conditions for 
the neutrino production and detection processes, which have to be 
fulfilled for these processes to be coherent, are always determined by the 
corresponding momentum uncertainties.

An interesting case in which the on-shell condition $\sigma_p\simeq \sigma_E$ 
is strongly violated both at production and detection is the proposed 
M\"ossbauer neutrino experiment, for which $\sigma_E\sim 10^{-11}$~eV and 
$\sigma_p\sim 10$~keV are expected \cite{AKL1,AKL2}, so that $\sigma_E\sim 
10^{-15} \sigma_p$. We discuss some important implications of this large 
disparity between $\sigma_E$ and $\sigma_p$ in Sec.~\ref{sec:sep}.

Our final comment in this section is on the case when $\sigma_p\sim\sigma_E
/v_g$ at neutrino production and detection, so that $\sigma_x\sim 
\sigma_p^{-1}$. While the localization condition requires 
relatively large $\sigma_p$ ($\sigma_p\gg \Delta m^2/2p$) for the emitted and 
detected neutrino states to be coherent superpositions of mass 
eigenstates (see Sec.~\ref{sec:coh1}), 
the condition of no decoherence due to the wave packet separation, 
on the contrary, requires long wave packets, i.e. relatively small $\sigma_p$. 
Is there any clash between these two requirements? By combining the two 
conditions we find $\Delta m^2/2p\ll \sigma_p\ll (v_g/\Delta v_g) L^{-1}$, 
which can only be satisfied if  
\be 
\Delta m^2/2p\ll  (v_g/\Delta v_g) L^{-1}\,.
\label{eq:consist1}
\ee  
This can be rewritten as the following condition on the baseline $L$: 
\be 
2\pi \frac{L}{l_{\rm osc}} \ll \frac{v_g}{\Delta v_g}\,.
\label{eq:consist2}
\ee  
Since $v_g/\Delta v_g\gg1$, this condition 
is expected to be satisfied with a large margin in any experiment which 
intends to detect neutrino oscillations: if it were violated, neutrino 
oscillations would have been averaged out because of the very large 
oscillation phase (except for unrealistically good experimental energy 
resolution $\delta E/E < \Delta v_g/v_g\sim \Delta m^2/2E^2)$.

\subsection{\label{sec:coh1}Coherence of the produced and 
detected neutrino states}

In order for a neutrino state produced in a charged-current weak interaction 
process to be a coherent superposition of different neutrino mass eigenstates, 
it should be in principle impossible to determine which mass eigenstate 
has been emitted. This means that the intrinsic quantum-mechanical 
uncertainty of the squared mass of the emitted neutrino 
state $\sigma_{m^2}$ must be larger than the 
difference $\Delta m^2$ of the squared masses of different neutrino mass  
eigenstates \cite{Kayser:1981ye,Rich:1993wu}: $\sigma_{m^2}\gtrsim \Delta 
m^2$. Conversely, if $\sigma_{m^2}\ll \Delta m^2$, one can determine which mass 
eigenstate has been emitted, i.e. the coherence of different mass eigenstates 
is destroyed. This situation is quite similar to that with the electron 
interference in double slit experiments:  If there is no way to find out which 
slit the detected electron has passed through, the detection probability will 
exhibit an interference pattern, but if such a determination is possible, the 
interference pattern will be washed out. 

Assume that by measuring the energies and momenta of the other particles 
involved in the production (or detection) process we can determine the energy 
$E$ and momentum $p$ of the emitted or absorbed neutrino state, and that 
the intrinsic quantum-mechanical uncertainties of these quantities are 
$\sigma_E$ and $\sigma_p$. From the energy -- momentum relation $E^2=p^2+m^2$ 
we can then infer the squared mass of the neutrino state with the uncertainty 
$\sigma_{m^2}=[(2 E \sigma_E)^2+(2 p\sigma_p)^2]^{1/2}$, where it is assumed 
that $\sigma_E$ and $\sigma_p$ are uncorrelated. Therefore the condition that 
the neutrino state be emitted or absorbed as a coherent superposition 
of different mass eigenstates is \cite{Kayser:1981ye,Rich:1993wu}
\be 
\sigma_{m^2}\equiv\left[(2 E \sigma_E)^2+(2 p\sigma_p)^2\right]^{1/2}\gg 
\Delta m^2\,. 
\label{eq:cond1} 
\ee 
This condition has a simple physical meaning. 
As follows from eq.~(\ref{eq:ineq1}), for relativistic neutrinos the 
second term in the square brackets in eq.~(\ref{eq:cond1}) exceeds the 
first one, and therefore this condition essentially reduces to 
$2p\sigma_p\gg \Delta m^2$, or 
\be
\sigma_p\gg \Delta m^2/2p=2\pi/l_{\rm osc}\,.
\label{eq:newloc}
\ee 
This is the so-called localization condition for neutrino production and 
detection. Indeed, since the effective production/detection momentum 
uncertainty $\sigma_p$ is related to the coordinate uncertainties of the 
neutrino emitter and absorber $\sigma_S^{\rm loc}$ and $\sigma_D^{\rm loc}$ 
by \\ 
$\sigma_p\sim min\{(\sigma_S^{\rm loc})^{-1},(\sigma_D^{\rm loc})^{-1}\}$, 
eq.~(\ref{eq:newloc}) yields  
\be 
\sigma_S^{\rm loc}\ll l_{\rm osc}\,,\qquad 
\sigma_{D}^{\rm loc}\ll l_{\rm osc}\,.
\label{eq:cond2} 
\ee 
This is nothing but the obvious requirement that the neutrino production 
and detection processes be localized in spatial regions that are small 
compared to the neutrino oscillation length; if it is violated, neutrino 
oscillations will be averaged out upon the integration over the neutrino 
production and detection coordinates in, respectively, the neutrino emitter  
and absorber. Such a washout of the oscillations is equivalent to decoherence. 

In real situations one always deals with large ensembles of neutrino emitters, 
and  the detectors also consist of a large number of particles. Therefore, in 
calculating the observable quantities -- the numbers of the neutrino detection 
events -- one always has to integrate over the macroscopic volumes of the 
neutrino source and detector.
In some situations (e.g., for solar or reactor neutrinos) the source 
and detector are much larger than the localization domains of the wave 
functions of individual neutrino emitters and absorbers. In these cases   
the integration over the source and detector volumes modifies the localization 
conditions: instead of depending on the spatial sizes of the individual 
neutrino emitter and absorber, $\sigma_S^{\rm loc}$ and $\sigma_D^{\rm loc}$, 
they contain the macroscopic lengths of the source and detector in the 
direction of the neutrino beam, $L_S$, and $L_D$. In other words, a necessary 
condition for the observability of neutrino oscillations is 
\be
L_{S,D}\ll l_{\rm osc}\,,
\ee
which can be much more restrictive than the conditions in eq.~(\ref{eq:cond2}). 
If this condition is violated, neutrino oscillations are averaged out.

Let us now discuss the coherence condition for the neutrino production and 
detection processes from a slightly different perspective. Quite generally, 
this condition can be formulated as a requirement that the energy and momentum 
uncertainties inherent in the neutrino production and detection processes be 
much larger than, correspondingly, the energy and momentum differences of 
different mass eigenstates $\Delta E$ and $\Delta p$:
\be
|\Delta E|\ll \sigma_E\,,\qquad |\Delta p|\ll \sigma_p\,. 
\label{eq:cond4}
\ee
As mentioned above, $\sigma_E$ is the effective energy uncertainty, which 
depends on the energy uncertainties both at production and detection and is 
dominated by the smallest between them, and similarly for the momentum 
uncertainty $\sigma_p$. It is now easy to see that if both the conditions in 
eq.~(\ref{eq:cond4}) are fulfilled, eq.~(\ref{eq:cond1}) is satisfied as 
well. %
\footnote{Indeed, by making use of the relation $\Delta m^2 \simeq 
2E\Delta E- 2p\Delta p$ (which follows from $m^2=E^2-p^2$ and is equivalent 
to (\ref{eq:exp1})), one can rewrite (\ref{eq:cond1}) as 
$\left[(2 E \sigma_E)^2+(2 p\sigma_p)^2\right]^{1/2}\gg
|2E\Delta E-2p\Delta p|$. From this representation of (\ref{eq:cond1}) 
the above statement is evident.}

Let us now find the relationship between the conditions (\ref{eq:cond4}) and 
the one in eq.~(\ref{eq:loc1}), which we called the interaction  
coherence condition in Sec.~\ref{sec:Iik}. As was pointed out in the previous 
subsection, the length of the wave packet $\sigma_x$ is determined by the 
effective energy uncertainty $\sigma_E$: $\sigma_x\simeq v_g/\sigma_E$.
Therefore one can rewrite the condition (\ref{eq:loc1}) as 
\be
\frac{|\Delta E|}{\sigma_E} \ll 1\,.
\label{eq:cond5}
\ee
This is nothing but the first condition in (\ref{eq:cond4}). To study 
the condition (\ref{eq:loc1}) further, let us rewrite (\ref{eq:cond5}), by 
making use of eq.~(\ref{eq:exp1}), as  
\be
\left|\frac{v_g\Delta p}{\sigma_E}\,+\,\frac{\Delta m^2}{2E\sigma_E}
\right|\,\ll\, 1\,.
\label{eq:cond6}
\ee
Now, there are essentially two possibilities:\\ 
(A) $v_g \Delta p$ is not approximately equal to $-\Delta m^2/2E$, and \\
(B) $v_g\Delta p\simeq -\Delta m^2/2E$. \\
Let us now consider these cases in turn.

(A). In this case there are no cancellations between the two terms on the 
l.h.s. of eq.~(\ref{eq:cond6}). From kinematics considerations it follows that 
case A is realized when the energy release at neutrino production is 
comparable to (or much larger than) the masses of the accompanying particles, 
as e.g. in the decays $\pi^+\to \mu^+ +\nu_\mu$ and $\pi^+\to e^+ 
+\nu_e$. In this case $\Delta E \sim \Delta p$. In the absence of 
cancellations between the two terms on the l.h.s. of (\ref{eq:cond6}) 
the condition (\ref{eq:cond6}) implies 
\be
\frac{v_g|\Delta p|}{\sigma_E}\ll 1\,,\qquad\quad \frac{\Delta m^2}
{2E\sigma_E}\ll 1\,.
\label{eq:cond7}
\ee
Since $\sigma_p > \sigma_E$, from the first strong inequality in 
(\ref{eq:cond7}) we find that for relativistic neutrinos $|\Delta p|/
\sigma_p\ll 1$, which is the second condition in (\ref{eq:cond4}). Thus, in 
case A the condition (\ref{eq:loc1}) enforces those in eq.~(\ref{eq:cond4}). 
 
Because $\sigma_E\simeq v_g/\sigma_x$ and $l_{\rm osc}=4\pi p/\Delta m^2$, 
the second condition in (\ref{eq:cond7}) is equivalent to 
$\sigma_x \ll l_{\rm osc}$, i.e. in this case the length of the 
neutrino wave packet is small compared to the oscillation length.

(B). Case B corresponds to a strong cancellation between the two terms on 
the l.h.s. of eq.~({\ref{eq:cond6}) (and between the two terms on the r.h.s. 
of eq.~(\ref{eq:exp1})). From eq.~(\ref{eq:exp1}) it follows that in this 
case $|\Delta E| \ll|\Delta p|$, which is close to the ``same energy'' 
situation. Case B is realized when the energy release in the 
neutrino production process is small compared to the masses of some of the 
accompanying particles; examples are nuclear $\beta$ decay and the decay 
$\pi^+\to \pi^0+e^+ +\nu_e$. Case B can be further subdivided into two cases:

\begin{itemize}

\item 
Case B$_1$: both terms on the l.h.s. of (\ref{eq:cond6}) are small, 
i.e. the conditions (\ref{eq:cond7}) are satisfied. The situation with 
coherence of neutrino production and detection in this case is the same as 
in case A, and therefore case B$_1$ does not require a separate analysis.  

\item 
Case B$_2$: both terms on the l.h.s. of (\ref{eq:cond6}) are not small, i.e.
\be
\frac{v_g |\Delta p|}{\sigma_E}\gtrsim 1\,,\qquad\quad \frac{\Delta m^2}
{2E\sigma_E}\gtrsim 1\,,
\label{eq:cond8}
\ee
and only their sum is small. Note that the signs $\gtrsim$ here can even be 
replaced by $\gg$ provided that the cancellation between the two terms in 
(\ref{eq:cond6}) is almost exact. Since $\sigma_E< \sigma_p$ (and may also 
be $\ll \sigma_p$), the first condition in (\ref{eq:cond8}) does not 
necessarily mean that the momentum differences of different mass eigenstates 
exceed the momentum uncertainties at production or detection, i.e. in general 
it does not contradict the second condition in (\ref{eq:cond4}); on the other 
hand, the condition $|\Delta p| \ll \sigma_p$ is not in this case 
automatically enforced by (\ref{eq:loc1}). 

The second inequality in (\ref{eq:cond8}) implies that the length 
of the neutrino wave packet is not small compared to the oscillation length. 
This is the case, in particular, for the proposed M\"ossbauer neutrino 
experiments \cite{AKL1,AKL2}. Note that the fact that the length of the wave 
packet can in some cases be larger (or even much larger) than the oscillation 
length does not mean that the oscillations cannot be observed in those cases. 
Indeed, from eq.~(\ref{eq:phase3}) it follows that, for a fixed $L$, the 
variation of the phase difference $\Delta \phi$ due to the variation of the 
quantity $|L-v_{g}t|$ between 0 and $\sigma_x$ does not exceed 
$(\Delta E/v_g)\sigma_x$. The condition (\ref{eq:cond5}) (which is equivalent 
to the one in eq.~(\ref{eq:loc1})) ensures that this variation is negligible. 
This situation actually corresponds to a quasi-stationary case, when the 
oscillatory pattern depends on the spatial coordinate but practically does not 
change with time. Therefore, if the sizes of the neutrino source and detector 
are small compared to the oscillation length, non-averaged oscillations will 
be observed despite the very large length of the wave packet.
\end{itemize}

Thus, we conclude that, in the wave packet approach, in cases A and B$_1$ the 
condition (\ref{eq:loc1}) guarantees the coherence of the neutrino production 
and detection processes. At the same time, in case B$_2$ characterized by 
eq.~(\ref{eq:cond8}), the condition (\ref{eq:loc1}) is not sufficient to 
ensure the coherence of neutrino production and detection and one has to 
supplement it by the second condition in (\ref{eq:cond4}), which in this case 
does not automatically follow from the wave packet formalism. Note that 
a strong cancellation between the two terms in (\ref{eq:cond6}) implies 
$|\Delta p|\simeq \Delta m^2/2p$, and therefore the second condition in 
(\ref{eq:cond4}) 
coincides in this case with the localization condition (\ref{eq:newloc})).

The above observations underline an important difference between the the 
wave packet approach and the more consistent quantum field theoretical
(QFT) treatment of neutrino oscillations, in which neutrino production,
propagation and detection are considered as a single process (see, e.g., 
\cite{Kobzarev:1980nk,Kobzarev:1981ra,Giunti:1993se,Rich:1993wu,Grimus:1996av,
Grimus:1998uh,Ioannisian:1998ch,Cardall:1999ze,Dolgov:1999sp,Beuthe2,
Dolgov:2002wy,Dolgov:2004ut,Dolgov:2005nb,AKL1}). 
The wave packet approach does not take into account the neutrino production
and detection processes, except by assigning to the neutrino state a momentum
uncertainty, which is supposed to be determined by these processes. In 
particular, the wave packet picture assumes the mass-eigenstate components of
the flavour neutrino states to be always on the mass shell, so that their
energy and momentum uncertainties are always related by eq.~(\ref{eq:rel1}).
As discussed above, the wave packet picture is quite adequate in cases 
A and B$_1$, but it needs to be supplemented by the information on the source 
and detector localization in case B$_2$. At the same time, in the QFT-based 
treatment of neutrino oscillations, a factor that takes into account the 
neutrino emitter and absorber localization properties emerges automatically 
in the expression for the oscillation probability in all cases.

This being said, we should add that, although case B discussed above is by 
no means rare, we are currently aware of only one instance when the situation 
in that case corresponds to the subcase B$_2$, i.e. is characterized by the 
conditions (\ref{eq:cond8}): this is the proposed M\"ossbauer neutrino 
experiment. To the best of our knowledge, for neutrinos from conventional 
sources the fulfillment of (\ref{eq:loc1}) always guarantees the coherence 
of the neutrino production and detection processes. It is for this reason that 
we dubbed it the interaction coherence condition.

\subsection{\label{sec:sep}Wave packet separation and restoration of coherence 
at detection}

Let us now assume that a neutrino flavour state was produced coherently 
in a weak interaction process and consider its propagation. The wave 
packet describing a flavour state is a superposition of the wave 
packets corresponding to different mass eigenstates. Since the latter 
propagate with different group velocities, after some time $t_{\rm coh}$ 
they will separate in space and will no longer overlap. If the 
spatial length of the wave packet is $\sigma_x$, this time is $t_{\rm coh}
\simeq \sigma_x/\Delta v_g$. The distance $l_{\rm coh}$ that the neutrino 
state travels during this time is $l_{\rm coh}\simeq v_g(\sigma_x/\Delta v_g) 
$. If the distance $L$ between the neutrino emission and detection 
points is small compared to the coherence length, i.e. if the condition 
(\ref{eq:coh1a}) is satisfied, then the coherence of the neutrino state 
is preserved, and neutrino oscillations can be observed. If, on the 
contrary, $L\sim l_{\rm coh}$ or $L\gg l_{\rm coh}$, partial or full 
decoherence should take place.
   
If the wave packet length $\sigma_x$ in the above discussion is assumed 
to be fully determined by the emission process, then this is not the full 
story yet: even if the coherence is lost on the way from the source to the 
detector, it still may be restored in the detector if the detection process 
is characterized by high enough energy 
resolution \cite{Kiers}. According to the quantum-mechanical time-energy 
uncertainty relation, high energy resolution requires the detection process 
to last sufficiently long; in that case, the different wave packets may arrive 
during the detection time interval and interfere in the detector (or, more 
precisely, the effects produced in the detector by different wave packets 
interfere). One can take into account this possible restoration of coherence 
in the detector by considering $\sigma_x$ to be an effective length of the 
wave packet, which exceeds the length of the emitted wave packet and takes 
the coherence of the detection process into account. This effective wave 
packet length is actually the length that characterizes the shape factors 
$G_i(L-v_{gi}t)$, as discussed in Secs.~\ref{sec:wp} and \ref{sec:size}. 
With $\sigma_x$ being the effective wave packet length, the coherence 
condition (\ref{eq:coh1a}) includes the effects of possible restoration of 
coherence at detection.  

An interesting example of coherence restoration at detection, which resolves 
another paradox of neutrino oscillations, is the proposed M\"ossbauer 
neutrino experiment. In this case $\sigma_E\sim 10^{-11}$~eV and $\sigma_p\sim 
10$~keV are expected \cite{AKL1,AKL2}. In such an  experiment neutrinos are 
produced coherently due to their large momentum uncertainty \cite{AKL1}. 
However, as soon as the emitted 
neutrino goes on the mass shell, its momentum uncertainty shrinks to satisfy 
eq.~(\ref{eq:rel1}), i.e. essentially becomes equal to the tiny energy 
uncertainty. Therefore for on-shell M\"ossbauer neutrinos $\sigma_p^{\rm eff}
\sim 10^{-11}$~eV~$\ll \Delta m^2/2E\simeq 10^{-7}$ eV, i.e. the momentum 
uncertainty is much smaller than the difference of the momenta of 
different mass eigenstates. This means that coherence of different mass 
eigenstates in momentum space is lost. However, the fact that in this case 
both the energy and momentum uncertainties of the propagating neutrino state 
are much smaller than $\Delta m^2/2E$ does not mean that oscillations cannot 
be observed. In fact, it has been shown in \cite{AKL1} that the M\"ossbauer 
neutrinos should exhibit the usual oscillations. The resolution of the paradox 
lies in the detection process: the large momentum uncertainty at detection, 
$\sigma_{p}\sim 10$ keV~$\gg \Delta m^2/2E$, restores the coherence by 
allowing the different mass eigenstates composing the flavour neutrino state 
to be absorbed coherently. 

It is well known that coherence plays a crucial role in observability of 
neutrino oscillations. It is interesting to note, however, that even 
non-observation of neutrino oscillations at baselines that are much shorter 
than the oscillation length is a consequence of and a firm evidence for 
coherence of the neutrino emission and detection processes: if it were 
broken (i.e. if the different neutrino mass eigenstates were emitted and 
absorbed incoherently), the survival probability of neutrinos of a given 
flavour, instead of being practically equal to 1, would correspond to averaged 
neutrino oscillations.

\section{\label{sec:stat}When is the stationary source approximation 
justified?}

It has been pointed out in \cite{Rich:1993wu,Kiers} and elaborated 
and exploited in \cite{Stodolsky:1998tc} that for stationary neutrino sources 
the following two situations are physically indistinguishable: 
\begin{itemize}
\item[(a)]
A beam of plane-wave neutrinos, each with a definite energy $E$ 
and with an overall energy spectrum $\Phi(E)$;

\item[(b)]
A beam of neutrinos represented by wave packets, each of them having the 
energy distribution function $f(E)$ such that $|f(E)|^2=\Phi(E)$.
 
\end{itemize}
As was stressed in \cite{Stodolsky:1998tc}, this actually follows from the 
fact that in stationary situations the spectrum $\Phi(E)$ fully determines 
the neutrino density matrix and therefore contains the complete information 
on the neutrino system. 

This, in fact, gives an alternative explanation of why the ``same energy'' 
approach, though based on an incorrect assumption, leads to the correct 
result. It has been shown in Sec.~\ref{sec:another} that, within the proper 
wave packet formalism, one can choose to sum up first the states of different 
mass but the same energy, and then integrate over the energy (or momentum)  
distributions described by the effective energy or momentum shape factors 
of the wave packets $h(E)$ or $h(p)$. In the light of the physical equivalence 
of the situations $(a)$ and $(b)$, it is obvious that the integration over the 
spectrum of neutrinos, which is inherent in any calculation of the event 
numbers, leads to the same result as the integration over the energy spread 
within the wave packets (provided that the corresponding energy distributions 
coincide). Thus, the ``same energy'' assumption, though by itself incorrect, 
leads to the correct number of events upon the integration over the neutrino 
energy spectrum. The same is true for the ``same momentum'' assumption. This 
actually means that the wave packet description becomes unnecessary in 
stationary situations, when the temporal structure of the neutrino emission 
and detection processes is irrelevant and the complete information on 
neutrinos is contained in their spectrum $\Phi(E)$, as was first pointed 
out in \cite{Stodolsky:1998tc}.

Let us derive the results of stationary source approximation in terms of 
the wave packet picture described in this paper. We start with 
eq.~(\ref{eq:ampl-p}) for the oscillation amplitude with $h_i(p)$ defined 
in (\ref{eq:h1}). Notice that the momentum distribution functions 
$f_i^S(p -  p_i)$ and 
$f_i^D(p- p_i')$ introduced in eqs.~(\ref{eq:wp1}) and (\ref{eq:wp1D}) 
do not depend on time, and furthermore they are defined for all moments $t$ 
from $- \infty$ to $+ \infty$. The only time dependence in (\ref{eq:wp1}) is 
in the form of the plane waves in the integrand. This is precisely what 
corresponds to the stationarity condition: the source has no special time 
feature, and there is no tagging of neutrino emission and detection times.

{}From eqs.~(\ref{eq:ampl-p}) and (\ref{eq:h1}) we find 
\bea
|{\cal A}_{ab}(L, t)|^2 = \sum_{i,k} U^*_{ai} U_{bi} U_{ak} U^*_{bk} 
\int d p \int d p^{\prime} f_i^S(p - p_i) f_k^{S*}(p^\prime - p_k) 
f^{D*}_i (p-p_i') f^{D}_k (p'-p_k') 
\hspace*{-5mm}
\nonumber \\
\times e^{i (p - p') L} 
e^{- i [E_i(p) -  E_k(p^\prime)]t}. 
\label{prob-ab}
\eea
The integration over time is trivial:  
\be
\int_{-\infty}^{+ \infty}  dt~ e^{- i [E_i(p) - E_k(p^\prime)]t} =  
2\pi \delta [E_i(p) - E_k(p^\prime)]\,,
\label{eq:int-t}
\ee
which means that only the waves with equal energies interfere. We stress once 
again that this is a consequence of the fact that no time structure appears 
in the detection and production processes, which is reflected in the time 
independence of the momentum distribution functions $f_i^S(p-p_i)$ and 
$f_i^D(p-p_i')$, and in the integration over the infinite interval of time.
The equality $E_i(p) = E_k(p^\prime)$ leads to 
\be
p - p^\prime = \Delta m_{ki}^2 /2p   
\ee
to the leading order in the momentum difference. The $\delta$-function 
(\ref{eq:int-t}) can be used to remove one of the momentum integrations  
in (\ref{prob-ab}), so that we finally obtain 
\be
P_{ab}(L)=
\int_{-\infty}^{+ \infty} dt |{\cal A}_{ab}(L, t)|^2 =  \frac{2\pi}{v_g}
\int d p  |f^S(p - \bar{p})|^2 |f^D (p-\bar{p}')|^2  
\sum_{i,k} U^*_{ai} U_{bi} U^*_{bk} U_{ak} 
e^{-i\frac{\Delta m_{ik}^2}{2p} L}. 
\label{eq:prob-int}
\ee
Here we have neglected the dependence of the shape factors on the neutrino 
mass. Replacing the integration over momenta by the integration over energies, 
we can rewrite the oscillation probability as 
\be
P_{ab}(L)=
\int_{-\infty}^{+ \infty} dt\, |{\cal A}_{ab}(L, t)|^2 = 
\frac{2\pi}{v_g^2} \int dE\, \Phi (E)\, R(E)\, P_{ab}(E,L), 
\label{eq:prob-int2}
\ee
where 
\be
P_{ab}(E,L) = \sum_{i,k} U^*_{ai} U_{bi} U^*_{bk} U_{ak}
e^{-i\frac{\Delta m_{ik}^2}{2E} L}
\label{prob-f}
\ee
is the standard expression for the oscillation probability, $\Phi(E) \equiv 
|f^S(E-\bar{E})|^2$ is the energy spectrum of the source, and $R(E) \equiv 
|f^D(E-\bar{E}')|^2$ is the resolution function of the detector (note that we 
have substituted the momentum dependence of these quantities by the energy 
dependence using the standard on-shell dispersion relation).

The expression in eq.~(\ref{eq:prob-int2}) corresponds to the stationary 
source approximation: the oscillation probability is calculated as an 
incoherent sum of the oscillation probabilities, computed for the 
same-energy plane waves, over all energies.  

Notice that we have performed integration over the spatial coordinate at 
the level of the amplitude and over time at the probability level. Apparently, 
such an asymmetry of space and time integrations is not justified from the 
QFT point of view. In QFT computations the integration over time is performed 
in the amplitude, and this leads (in the standard setup) to the delta function 
which expresses the conservation of energy in the interaction process. To 
match our picture with that of QFT we need to consider the detection process 
and take into account the energies of all the particles that participate in 
the process. Suppose that the algebraic sum of the energies of all the 
accompanying particles (taken with the ``-'' sign for all incoming 
particles the ``+'' sign for the outgoing ones) is $E_D$. Then instead of 
(\ref{eq:int-t}) we will have in the probability
\be
\int_{-\infty}^{+ \infty} dt\, e^{- i [E_i(p) - E_D]t}
\int_{-\infty}^{+ \infty} dt'\, e^{i [E_k(p') - E_D]t'}
= 4\pi^2\,\delta[E_i(p)- E_k(p')]\,\delta[E_i(p) - E_D]\,, 
\label{eq:int-t2}
\ee
where the second $\delta$-function on the right hand side reflects the energy 
conservation in the detection process. Using (\ref{eq:int-t2}) we again obtain 
the ``same energy'' interference, as before.

In the above calculation we have not introduced any time structure at 
detection and performed the integration over $t$ from $-\infty$ to $+\infty$. 
In reality, certain time scales are always involved in the detection processes 
(even if we do not perform any time tagging in the emission process). For 
example, we can measure with some accuracy the appearance time of a charged 
lepton produced by the neutrino capture in the detection process. In this 
case, the neutrino detection state will have a time dependence: 
\be
\Psi_i^D  = \Psi_i^D (x - L, t - t_0)\,, 
\ee
where $\Psi_i^D$ has a peak at $t_0$ of a width $\sigma_t$ that is determined 
by the accuracy of the measurement of the time of neutrino detection. Since 
in practice the spatial characteristics of neutrino detection do not change 
with time, the dependences of $\Psi_i^D$ on $x$ and $t$ factorize: $\Psi_i^D 
= \Psi_{x i}^D(x - L) \Psi_{t i}^D (t - t_0)$. Integrating over time in the 
amplitude, we will have 
\be
\int dx \int dt \,  \Psi_i^{D*} (x - L, t - t_0)\,e^{ipx - i (E_i-E_D) 
t} = 2\pi\,
e^{ip L} f_{i}^{D*}(p-p_i')\, f_{t i}^{D*}(E_i-E_D)\,,
\ee
where 
\be
f_{t i}^D (E) \equiv \int\frac{dt}{\sqrt{2\pi}}\,\Psi_{t i}^{D}(t - t_0)
\, e^{i E t} 
\ee
is the Fourier transform of $\Psi_{t i}^{D}(t-t_0)$. As we have mentioned, 
$\Psi_{t i}^D (t - t_0)$ has a peak of the width $\sigma_t$ at $t=t_0$. 
Taking for an estimate
$\sigma_t \sim 10^{-9}$ s (which is probably 
the best currently achievable time resolution), we obtain $\delta E 
\sim\sigma_t^{-1}\sim 10^{-6}$ eV. This is many orders of magnitude 
smaller than the typical energy resolution in the oscillation experiments. 
Therefore one can substitute $f_{t i}^D (E_i-E_D) \rightarrow \sqrt{2\pi}\,
\delta(E_i - E_D)$, which brings us back to our previous consideration.

\section{\label{sec:when}When can neutrino oscillations be described by 
production and detection independent probabilities?}

In most analyses of neutrino oscillations it is assumed that the oscillations 
can be described by universal, i.e. production and detection process 
independent probabilities. In other words, it is assumed that by specifying 
the flavour of the initially produced neutrino state, its energy and 
the distance between the neutrino source and detector, one fully determines the 
probability of finding neutrinos of all flavours at the detector site (for 
known neutrino mass squared differences and leptonic mixing matrix). 
The standard formula for neutrino oscillations in vacuum, eq.~(\ref{eq:P2}), 
is actually based on this assumption. Such an approach is very often well 
justified, but certainly not in all cases.  It is, therefore, interesting 
to study the applicability limits and the accuracy of this approximation.  

A natural framework for this is that of QFT, which provides the most 
consistent approach to neutrino oscillations. In this method the neutrino 
production, propagation and detection are considered as a single process with 
neutrinos in the intermediate state. This allows one to avoid any discussion 
of the properties of the neutrino wave packets since neutrinos are actually 
described by propagators rather than by wave functions. The properties of 
neutrinos in the intermediate state are fully determined by those of the 
``external'' particles, i.e. of all the other particles that are involved in 
the neutrino production and detection processes. The wave functions of these 
external particles have to be specified. Usually, these particles are assumed 
to be described by wave packets; for this reason the QFT-based treatment is 
often called the ``external wave packets'' approach \cite{nuwp2}, as opposed 
to the usual, or ``internal wave packets'' one, which was discussed in 
Secs.~\ref{sec:why}, \ref{sec:another} and \ref{sec:stat}   
and which does not include neutrino 
production and detection processes. The results of the QFT-based approach turn 
out to be similar, but not identical, to those of a simple wave packet one; in 
particular, possible violations of the on-shell relation (\ref{eq:rel1}) 
between the neutrino energy and momentum uncertainties is now automatically 
taken into account. Moreover, the values of these uncertainties, which specify 
the properties of the neutrino wave packet in the ``internal wave packets'' 
approach and which have to be estimated in that approach, are now directly 
derived from the properties of the external particles. 

The results of the QFT approach can be summarized as follows. For 
neutrinos propagating macroscopic distances the overall probability of 
the production--propagation--detection process for relativistic or 
quasi-degenerate neutrinos can to a very good accuracy be represented as 
a product of the individual probabilities of neutrino production, 
propagation (including oscillations) and detection.%
\footnote{A notable exception from this rule is the case of the M\"ossbauer 
effect with neutrinos, in which the probabilities of neutrino production and 
detection do not factorize but are instead entangled with each other 
\cite{AKL1}. Still, even in this case, the oscillation (actually, 
$\bar{\nu}_e$ survival) probability can to a very good accuracy be 
factored out of the expression for the overall probability of the 
process.}  
The oscillation probability, however, is not in general 
independent of production and detection processes, which means that the 
factorizability of the probability of the entire production - 
propagation - detection process and the universality of the oscillation 
probability (or lack thereof) are in general independent issues. The 
oscillation probability can be generically represented as \be 
P(\nu_a\to\nu_b;\,L) = \sum_{i,k} U^*_{ai} U_{bi} U_{ak} U_{bk}^*\, 
\,e^{-i\frac{\Delta m_{ik}^2}{2p} L}\,\,S_{\rm coh}(L/l^{\rm 
coh}_{ik})\,S_{\rm P/D}(\Delta m^2_{ik}/\sigma_{m^2})\,. \label{eq:P4} 
\ee Here $S_{\rm P/D}(\Delta m^2_{ik}/\sigma_{m^2})$ and $S_{\rm coh} 
(L/l^{\rm coh}_{ik})$ are, respectively, the production/detection and 
propagation coherence factors, which account for possible suppression of 
the oscillations due to the lack of coherence at neutrino production or 
detection and due to the wave packet separation. They are both equal to unity 
at zero argument and quickly decrease (typically exponentially) when their 
arguments become large. 
The simple ``internal wave packets'' approach leads to an expression for the 
oscillation probability that is similar in form to that in eq.~(\ref{eq:P4}) 
but with the argument of $S_{\rm P/D}$ replaced by $(\Delta E_{ik}/\sigma_E)$. 
As discussed in Sec.~\ref{sec:coh1}, the two production/detection coherence 
conditions,  $\Delta m^2_{ik}\ll \sigma_{m^2}$ and $\Delta E_{ik}\ll 
\sigma_E$, are essentially equivalent except when the conditions 
(\ref{eq:cond8}) are satisfied.  Note that in the QFT-based framework the 
form of the propagation and production/detection coherence factors 
$S_{\rm coh}(L/l^{\rm coh}_{ik})$ and $S_{\rm P/D}(\Delta m^2_{ik}/
\sigma_{m^2})$ in eq.~(\ref{eq:P4}) is derived from the properties of the 
external particles and of the detection and production processes.

For the oscillation probability to be independent of the processes of neutrino 
production and detection, the following conditions have to be fulfilled:

\begin{itemize}

\item[(i)]
The neutrino emission and absorption should be coherent, and decoherence 
effects due to wave packet separation should be negligible;

\item[(ii)]
The energy release in the production and detection reactions should be  
large compared to the neutrino mass (or compared to the mass differences). 

\end{itemize}
The necessity of (i) is clear from the discussion above: if this condition is 
fulfilled, both the coherence factors in eq.~(\ref{eq:P4}) are equal to unity, 
and the standard neutrino oscillation formula is recovered. If, on the 
contrary, (i) is violated, the oscillations will in general suffer from the 
production and detection dependent decoherence effects (note, however, that 
if the decoherence is complete, the oscillation probability will still have a 
universal form as it will correspond to averaged oscillations in that case). 
As to the condition (ii), it ensures that the production and detection 
probabilities are essentially the same for all mass-eigenstate components of 
the emitted or detected flavour neutrino states (modulo the different values 
of $|U_{ai}|^2$); if this condition is violated, the phase space available in 
the production or detection process will depend on the mass of the 
participating neutrino mass eigenstate, and the mass-eigenstate composition of 
the flavour eigenstates will no longer be given by the simple formula 
(\ref{eq:mix}).

\section{\label{sec:Lorentz}Wave packet approach and Lorentz invariance}

Let us start the discussion of the Lorentz invariance issues with considering 
yet another apparent paradox. In Sec.~\ref{sec:size} it was pointed out that 
the energy uncertainty of neutrinos produced in decays of free or quasi-free 
unstable particles at rest is essentially given by the parent particle's 
decay width, i.e. $\sigma_E\simeq \Gamma$. 
Correspondingly, the spatial length of the neutrino wave packet is 
\be
\sigma_x\simeq v_g/\sigma_E\simeq v_g/\Gamma=v_g\, \tau\,,
\label{eq:sigma0}
\ee
where $\tau$ is the mean lifetime of the decaying particle. 

Consider now the situation when in the laboratory frame the parent particle 
moves with a speed $u$. One might then expect that the neutrino energy 
uncertainty in the laboratory frame $\sigma_E'$ will be given by the decay 
width of the unstable particle in this frame $\Gamma'=\Gamma/\gamma_u$, where 
$\gamma_u=1/\sqrt{1-u^2}$ is the Lorentz factor of the parent particle. The 
length of the neutrino wave packet $\sigma_x'$ is then expected to be $v_g'/
\Gamma'\simeq 1/\Gamma'=\gamma_u/\Gamma$, which means that $\sigma_x'$ should 
be increased by a factor of $\gamma_u$ compared to $\sigma_x$. On the other 
hand, the transition from the rest frame of the unstable particle to the 
laboratory frame is described by a Lorentz boost; if the parent 
particle is boosted in the direction of the neutrino momentum, then the length 
of the neutrino wave packet should be reduced due to the Lorentz contraction, 
not increased! 

The resolution of this paradox comes from the observation that the decay 
process takes a finite time $\tau'=1/\Gamma'$, during which the decaying 
particle propagates some distance. Consider the situation when 
the neutrino is emitted along the momentum of the parent particle. 
During the decay time the parent particle moves over distance $l=u \tau'$ in 
the direction of the motion of the neutrino, thus reducing the length of  
the emitted neutrino wave packet:
\be
\sigma_x'\simeq v_g'/\Gamma' - l=v_g' \tau'-u \tau'= (v_g' -u)\gamma_u\tau=
\frac{v_g \tau}{\gamma_u(1+v_g u)}\,,
\label{eq:sigmax}
\ee 
where in the last equality we used the relativistic law of addition of 
velocities 
\be
v_g'=\frac{v_g+u}{1+v_g u}\,.
\label{eq:relat}
\ee
Comparing eq.~(\ref{eq:sigmax}) with eq.~(\ref{eq:sigma0}), we find
\be
\sigma_x'=\frac{\sigma_x}{\gamma_u(1+v_g u)}\,,
\label{eq:sigmax2}
\ee
which is just the standard relativistic law of transformation of lengths in 
the direction of the motion of objects.%
\footnote{Upon a boost with the speed $v$ from its rest frame to a frame $K$, 
the length of an object in the direction of the boost $d^0$ is contracted by  
the factor $\gamma_v=1/\sqrt{1-v^2}$: $d=d^0/\gamma_v$. In the reference frame 
$K'$ which moves in the same direction as $K$ with the speed $u$ with respect 
to the latter, one has $d'=d^0/\gamma_{v'}$, where $v'=(v+u)/(1+v u)$. From 
the latter equality one finds $\gamma_{v'}=\gamma_u \gamma_v (1+uv)$. This 
gives $d'=d \gamma_v/\gamma_{v'}=d/[\gamma_u (1+v u)]$, to be compared with 
eq.~(\ref{eq:sigmax2}).}
For relativistic neutrinos $v_g\simeq v_g'\simeq 1$ and  
\be
\gamma_u(1+v_g u)\simeq \gamma_u(1+u)=\sqrt{\frac{1+u}{1-u}}\,,
\label{eq:factor}
\ee
i.e. the neutrino wave packets are contracted when neutrinos are emitted in 
the direction of the motion of the parent particles and dilated when they are  
emitted in the opposite direction ($u<0$), as it should be. This result was  
previously obtained on the basis of different arguments in \cite{FarSm}.

Let us now discuss the behaviour of the oscillation probability with respect 
to the Lorentz boosts. We first analyze the standard expression for the 
oscillation probability (\ref{eq:P2}) (our discussion here mostly follows 
that in \cite{Levy}; a more general though less direct proof will be 
given below). Consider a boost with the speed 
$u$ of the neutrino source -- detector system in the direction of the neutrino 
motion (the generalization to the boost in an arbitrary direction is 
straightforward). Under this boost the distance $L$ traveled by the neutrino 
between its emission and absorption points, the propagation time $t$ and 
the neutrino energy and momentum transform according to
\bea
L'=\gamma_u(L+ut)\,,\,\qquad t'=\gamma_u(t+uL)\,,~
\label{eq:trans1} \\
E'=\gamma_u(E+up)\,,\!\qquad p'=\gamma_u(p+uE)\,.\,
\label{eq:trans2}
\eea
Recall now that the standard formula for the oscillation probability results 
when the production/detection and propagation coherence conditions are 
satisfied; for neutrinos from conventional sources the former condition 
includes the requirement that the neutrino wave packet length $\sigma_x$ be 
small compared to the oscillation length $l_{\rm osc}$ (M\"ossbauer neutrinos, 
for which this condition is not satisfied, represent a special case, to be 
considered below). Since the baselines of interest in neutrino oscillation 
experiments satisfy $L\gtrsim l_{\rm osc}$, the condition $\sigma_x\ll 
l_{\rm osc}$ also means that the length of the neutrino wave packets is 
negligible compared to the baseline, so that one can consider neutrinos 
pointlike and set $L=v_g t$. Substituting this into the first equality in 
eq.~(\ref{eq:trans1}) yields $L'=\gamma_u L(1+u/v_g)$. On the other hand, 
since $v_g=p/E$, the second equality in eq.~(\ref{eq:trans2}) can be rewritten 
as $p'=\gamma_u p(1+u/v_g)$. Thus 
\be
\frac{L'}{p'}\,=\,\frac{L}{p}\,,
\label{eq:rat1}
\ee
which proves Lorentz invariance of the standard oscillation probability 
(\ref{eq:P2}).

{}Note that the first equality in (\ref{eq:trans2}) can be written as  
$E'=E \gamma_u (1+u v_g)$, and therefore eq.~(\ref{eq:sigmax2}) implies  
that the product $\sigma_x E$ is invariant under Lorentz transformations. 
This result was previously obtained in \cite{FarSm} from different 
considerations. 

We shall now present an alternative (and more general) proof of Lorentz 
invariance of the standard oscillation probability. The phase difference 
$\Delta\phi$ defined in eq.~(\ref{eq:phase1}) is Lorentz invariant;  
hence so is the expression for $\Delta\phi$ in eq.~(\ref{eq:phase3}), which 
was obtained from (\ref{eq:phase1}) assuming neutrinos to be  relativistic or 
quasi-degenerate. (Recall that we adopt this assumption throughout this 
paper, and that the standard expression for the oscillation probability 
(\ref{eq:P2}) is only valid under this assumption).  
The two terms in eq.~(\ref{eq:phase3}) are not in general separately  
Lorentz invariant, though their sum is. However, in the situations when the 
second term is negligibly small in all Lorentz frames, the first term 
must be Lorentz invariant by itself; the standard oscillation probability 
(\ref{eq:P2}) is then Lorentz invariant. 
As discussed in Secs.~\ref{sec:Iik} and \ref{sec:coh1}, the second term 
in eq.~(\ref{eq:phase3}) is responsible for possible violation of the 
production/detection coherence condition, and can be discarded 
when this condition is satisfied. From eq.~(\ref{eq:phase3}) it follows that 
this term vanishes, in particular, in the limit of pointlike neutrinos, 
$L=v_g t$, which leads to Lorentz invariance of the standard oscillation 
probability, as discussed above. In addition, this term vanishes in the limit 
of vanishingly small $\Delta E$ even if the condition $L=v_g t$ is not 
satisfied; this case is realized for M\"ossbauer neutrinos \cite{AKL1,AKL2}), 
whose oscillation probability is therefore also Lorentz invariant. We 
add here that the condition $L=v_gt$ is Lorentz invariant, and so is 
the condition $\Delta E\simeq 0$ for M\"ossbauer neutrinos; the former 
follows directly from eq.~(\ref{eq:trans1}), while the latter is a 
consequence of the fact that M\"ossbauer neutrinos are emitted and 
absorbed recoillessly in any Lorentz frame. 

How about the general case when the production/detection and propagation 
coherence conditions are not necessarily satisfied? The oscillation 
probability must be Lorentz invariant in that case as well. In the QFT-based 
approach, this comes out automatically since the calculations are based on 
Lorentz covariant Feynman rules. In contrast to this, calculations in the wave 
packet approach are not manifestly Lorentz covariant; however, as we shall 
show now, they still lead to the Lorentz invariant expression for the 
oscillation probability.  

Let us first demonstrate that the integral over time of the product of any two 
wave packet functions, $G_i(L-v_{gi}t)$ and $G_k^*(L-v_{gk}t)$, is Lorentz 
invariant.
We write the effective wave packets as $G_i=|G_i|e^{i\rho_i}$. From the 
normalization condition (\ref{eq:norm}) it follows that the quantities 
$|G_i|^2 dt$ are Lorentz invariant for all $i$, i.e. 
\be
|G_i|^2 dt=inv\,, \qquad |G_k|^2 dt=inv\,. 
\ee
Taking the square roots of these two equalities and 
multiplying the results, we find 
\be
|G_i||G_k| dt=inv\,.  
\label{eq:inv}
\ee
Next, we consider the complex phase factors $e^{i\rho_i}$. They are Lorentz 
invariant because so must be the phases $\rho_i$. The latter is a consequence 
of the fact that for any physical quantity that can be expanded in a 
nontrivial power series the expansion parameter must be Lorentz invariant -- 
otherwise different terms in the expansion would transform differently under 
the Lorentz transformations. 
{}From the invariance of $e^{i\rho_i}$ and eq.~(\ref{eq:inv}) it 
follows that the integral over time of the product of any two wave packet 
functions, $G_i(L-v_{gi}t)$ and $G_k^*(L-v_{gk}t)$, is Lorentz invariant, 
as advertised.%
\footnote{Note that in general this integral is not an $L$-independent 
constant, unlike the normalization integral (\ref{eq:norm}). Still, it 
is Lorentz invariant. We encourage the reader to check that explicitly 
for Gaussian wave packets (see Appendix B).}
Since so is the phase difference $\Delta\phi$, the quantities $I_{ik}(L)$ 
defined in eq.~(\ref{eq:Iik1}) are Lorentz invariant as well. Lorentz 
invariance of the oscillation probabilities $P_{ab}(L)$ then  
follows immediately from eq.~(\ref{eq:P3}).

Before closing the discussion of the Lorentz invariance issues, let us note 
an interesting feature of the wave packets $G_i(L-v_{gi}t)$. 
As discussed in Sec.~\ref{sec:wp}, their moduli 
decrease rapidly when $|L-v_{gi}t|$ exceeds 
the effective length of the wave packet $\sigma_x$. Therefore $|G_i|$ 
should depend on $(L-v_{gi}t)$ through the ratio 
\be 
R_i \equiv 
\frac{L-v_{gi}t}{\sigma_{xi}}\,. 
\label{eq:rat2} 
\ee 
Under the Lorentz transformations we have 
\be 
L'-v_{gi}'t'=\gamma_u\left[(L+u t)-\frac{v_{gi}+u}{1+v_{gi} u}(t+u L)
\right] =\frac{L-v_{gi}t}{\gamma_u (1+v_{gi} u)}\,. 
\label{eq:new} 
\ee 
{}From this formula and the properties of $\sigma_x$ under Lorentz 
transformations (\ref{eq:sigmax2}) it follows that the ratio (\ref{eq:rat2}) 
is Lorentz invariant. Note that, in addition to their dependence on $R_i$, 
$|G_i(L-v_{gi}t)|$ depend on $\sigma_{xi}$ and $v_{gi}$ through their 
normalization factors.

\section{\label{sec:remarks}Remarks on the wave packet approach}

We add here a few comments on the wave packet approach to 
neutrino oscillations. 

(1) In the derivation of the oscillation probability in the framework 
of the wave packet approach, the integration of the squared modulus of 
the transition amplitude over time is involved. This is usually motivated by 
the fact that in most experiments the neutrino emission and detection times
are not measured. One might therefore wonder if, when considering the 
experiments like K2K or MINOS in which the neutrino time of flight is 
actually measured, the integration over time is redundant or even 
illegitimate.  

In reality, the integration over time is necessary in any case due to the 
fact that the neutrino emission and detection processes are not instantaneous 
but rather take a finite time. This is reflected in the finite temporal
extension of the neutrino wave packets $\sigma_t\sim \sigma_E^{-1} \sim 
\sigma_x/v_g$ (note that $\sigma_t$ depends on both the production and 
detection processes).  One has to remember that even in the experiments 
like K2K or MINOS the neutrino production and detection are not instantaneous 
but always take place within certain time windows. 
Consider, e.g. the K2K experiment \cite{K2K}. The neutrinos were produced in 
the decays of pions originating from 1.1 $\mu$s proton spills, and the time 
window for the neutrino detection was 1.5 $\mu$s. These time intervals are 
much shorter than the neutrino time of flight ($\sim 833~\mu$s), but much 
longer than the time extension of the neutrino wave packets 
$\sigma_t\sim 10^{-3}~\mu$s.%
\footnote{The length of the neutrino wave packet in the rest frame of 
the parent pion is essentially given by the pion lifetime: $\sigma_x^0\sim 
c\tau_\pi=780$ cm. In the laboratory frame it is about 1.5 orders of magnitude 
smaller due to the Lorentz contraction, which yields the above estimate 
for $\sigma_t\sim \sigma_x/v_g$ (assuming that the time scales of the neutrino 
production and detection processes are of the same order of magnitude).} 
Thus, the time windows in the K2K 
experiments exceeded significantly the characteristic times of neutrino 
emission and detection, which justifies the infinite-limits integration 
over time.

(2) It was pointed out in Sec.~\ref{sec:wp} and discussed in Sec.~\ref{sec:Iik} 
that in general the mean momenta of the wave packets describing the neutrino 
production and detection states need not coincide, ${\bf p}_i \ne {\bf p}_i'$. 
This comes about because the conditions of neutrino production and detection 
are in general different. They are determined by the features of these 
processes (e.g., by the energy dependence of the production and detection 
amplitudes, the energy thresholds, the selection of certain energy intervals 
of the accompanying particles, etc.). 

At the same time, in the QFT-based approach to neutrino oscillations the 
neutrino is in the intermediate state and its evolution is described by a 
propagator characterized, in the momentum space, by a single momentum. 
The correspondence between the wave packet and QFT pictures can be established 
by noticing that in the wave packet formalism the functions $G_i$ play to some 
extent the same role as the propagators in the QFT approach (see below), 
and that there are two different representations for these functions. These 
representations, given in eqs.~(\ref{eq:G}) and (\ref{eq:G1}), correspond 
to two different orders of integrations over the momenta of the production and 
detection states on the one hand and projection of the evolved produced state 
onto the detected state, which involves an integration over the spatial 
coordinate, on the other hand. In one case we first integrate over the momenta 
of the produced and detected neutrino states, thus forming the wave packets 
with certain mean momenta ${\bf p}_i$ and ${\bf p}_i'$, and then project the 
evolved produced state onto the detected state, as in eq. (\ref{eq:G}). The 
difference of the mean momenta then appears through the factor 
$e^{i ({\bf p}_i - {\bf p}_i')({\bf x} - {\bf L})}$ in the integrand. This 
factor suppresses the projection if the average momenta are significantly 
different. In the other picture, which is illustrated by eq.~(\ref{eq:G1}), 
one considers the projection of the plane waves composing the production and 
detection wave packets and then integrates over the common momentum $p$. This 
common for the produced and detected states momentum (the same for $f_i^S$ and 
$f_i^D$) appears as a result of the integration over the spatial coordinate $x$ 
involved in the projection procedure, which leads to a delta function that 
eliminates one of the momentum integrations. The difference of the mean 
momenta $\delta_i=p_i-p_i'$ appears then as a relative shift of the arguments 
of the shape factors in the momentum space.

This second picture with a single integration over the momentum corresponds 
more closely to the one in the QFT approach. As we have already mentioned, the 
function $G_i$ is a wave-packet analogue of the neutrino propagator of the QFT 
approach, which can be clearly seen from the representation of $G_i$ 
in eq.~(\ref{eq:G1}). Unlike the usual propagator, however, it contains, 
in addition, certain information about the neutrino production and detection 
processes.

(3) In this paper we have concentrated on neutrino oscillations in vacuum. 
In this case the wave packet approach (which implies a subdivision of the 
whole process into three stages: production, propagation and detection) and 
the QFT-based approach, which treats all three stages jointly, have 
comparable complexity, and the correspondence between them can be relatively 
easily established. In the case of neutrino oscillation in medium with a 
non-trivial density profile, the partition of the overall processes 
substantially simplifies the picture. As an example, solar neutrinos undergo 
flavor conversion inside the Sun, than propagate between the Sun and the Earth 
and then oscillate inside the Earth; the evolution of neutrinos has a rather 
complicated character, even when neutrinos are treated as being strictly 
real (on-shell) particles. In this case using the wave packet approach is 
essential, and it is important to understand its domain of applicability and 
limitations.

(4) The coherent propagation condition (\ref{eq:coh1a}) is the requirement 
that the wave packet separation upon propagation over the distance $L$ be 
small compared to the effective spatial length of the wave packet $\sigma_x$. 
It can be reformulated in the momentum space as the requirement that the 
variation of the oscillation phase within the wave packet due to the momentum 
(or energy) spread be small. Indeed, since $\Delta\phi_{st}=(\Delta m^2/2p)L$, 
the latter condition gives
\be
\left|\frac{\partial\Delta\phi_{st}}{\partial p}\frac{\partial p}{\partial E}
\right|\sigma_E = \frac{\Delta m^2}{2 p^2}L\,\frac{\sigma_E}{v_g} \ll 
1\,,
\label{eq:coh1b}
\ee
which is equivalent to (\ref{eq:coh1a}). 

At the same time, the interaction coherence condition  
(\ref{eq:loc1}), which requires the variation of the oscillation phase along 
the wave packet in the coordinate space to be small, translates in the 
momentum space into the condition of no (or little) separation in the momentum 
variable between the wave packets describing different mass eigenstates.  

Mathematically, the loss of coherence due to the spatial separation of the 
wave packets is reflected in the lack of overlap of the factors $G_i$ and 
$G_k$ in the integrand of the coordinate-space integral representation 
(\ref{eq:Iik2}) for $I_{ik}(L)$, and in fast oscillations of the integrand in 
the momentum-space integral representation (\ref{eq:I3}). For the 
interaction coherence condition the situation is converse: its 
violation is reflected in fast oscillations of the integrand in the 
coordinate-space integral (\ref{eq:Iik2}) and in the lack of overlap of the 
$f_i^{S,D}$ and $f_k^{S,D}$ factors in the integrand of the momentum-space 
integral (\ref{eq:I3}). 
Thus, there is certain duality between the wave packet descriptions of 
neutrino oscillations in the coordinate space and in the momentum space.

\section{\label{sec:disc}Discussion and summary}

In the present paper we discussed a number of subtle issues of the 
theory of neutrino oscillations which are still currently under debate 
or have not been sufficiently studied yet. For each issue, we were 
trying to present our analysis from different perspectives and obtained 
consistent results. We have developed a general wave packet approach to 
neutrino oscillations, which does not rely on the specific form of 
the wave packets. Our approach is formulated in terms of the 
generalized wave packets whose effective characteristics depend on the 
properties of the neutrino production and detection processes. 
In particular, 
we took into account that the momentum distribution functions of the produced 
and detected neutrino states are in general different, and explored physical 
consequences of this difference. 

We have also developed a new approach to calculating the oscillation 
probability in the wave packet picture, in which we changed the usual order of 
integration over the momenta (or energies) and summation over the mass 
eigenstate components of the wave packets representing the flavour neutrino 
states. This allowed a new insight into the question why the 
generally unjustified 
``same energy'' and ``same momentum'' assumptions lead to the correct result 
for the oscillation probability.

We have also presented an alternative derivation of the equivalence between 
the results of the sharp-energy plane wave formalism and of the wave packet 
approach in the case of stationary neutrino sources, as well as discussed 
the applicability conditions for the stationary source approximation.

Below we give a short summary of our answers to the first seven 
questions listed in the Introduction.

(1) The standard formula for the probability of non-averaged neutrino 
oscillations is obtained if the neutrino production and detection processes 
are coherent and the decoherence effects due to the wave packet separation 
are negligible. Under these conditions the additional phase $\Delta \phi'$, 
which characterizes the phase variation along the neutrino wave packet,  
is negligible. The ``same energy'' and ``same momentum'' assumptions, 
which allow one to nullify this additional phase, are then unnecessary. These 
assumptions still lead to the correct result because their main effect is 
essentially just to remove this extra phase. 

An alternative explanation of the fact that the ``same energy'' assumption 
gives the correct result comes from the observation that in going from 
the oscillation probability to the observables such as event numbers, 
one has to integrate over the neutrino spectra. As discussed in 
\cite{Kiers,Stodolsky:1998tc} and in Sec.~\ref{sec:stat}, for stationary 
neutrino sources this is equivalent to integration over the energy 
distribution within wave packets (provided that this energy distribution 
coincides with the spectrum of plane-wave neutrinos). In 
Sec.~\ref{sec:another} we have shown that the integration over the spectrum 
of plane-wave neutrinos is just a calculational convention in the wave 
packet approach, which does not involve any additional approximations.

(2) Quantum-mechanical uncertainty relations are at the heart of the 
phenomenon of neutrino oscillations. For neutrino production and detection to 
be coherent, the energy and momentum uncertainties inherent in these 
processes must be large enough to prevent a determination of the neutrino's 
mass. These uncertainties are governed by the quantum-mechanical uncertainty 
relations, which also determine the size of the neutrino wave packets and 
therefore are pivotal for the issue of the coherence loss due to the wave 
packet separation. 

(3) The spatial size of the neutrino wave packets is always determined by 
their energy uncertainty $\sigma_E$. Note that $\sigma_E$ is the effective 
uncertainty which depends on the energy uncertainties both at neutrino 
production and detection.

(4) The coherence propagation condition ensures that the wave packets 
corresponding to the different mass eigenstates do not separate to such an 
extent that their effects can no longer interfere in the detector. This 
condition is therefore related to the spatial size of the neutrino wave 
packets, which is determined by the neutrino energy uncertainty $\sigma_E$. 

At the same time, the localization condition, which has to be satisfied in 
order for the neutrino production and detection processes to be coherent, is  
determined by the effective momentum uncertainty $\sigma_p$, which depends on 
the momentum uncertainties at production and detection. In the wave packet 
approach, the neutrino energy and momentum uncertainties are related to each 
other due to the on-shellness of the propagating neutrino, whereas in a more 
general quantum field theoretic framework they are in general unrelated.

(5) The wave packet approach (or a superior quantum field theoretic one) are 
necessary for a consistent derivation of the expression for the oscillation 
probability. Once this has been done, wave packets can be forgotten in all 
situations except when the decoherence effects due to the wave packet 
separation or due to the lack of localization of the neutrino source or 
detector become important. Even in those cases, though, the decoherence 
effects can in most situations be reliably estimated basing on the standard 
oscillation formula and simple physical considerations. 
The wave packet approach is, however, useful for analyzing a number of subtle 
issues of the theory of neutrino oscillations.  

(6) The oscillation probability is independent of the production and 
detection processes provided the following conditions are satisfied: $(i)$ 
neutrino emission and absorption are coherent, 
and decoherence effects due to the wave packet separation 
are negligible, and $(ii)$ the energy 
release in the production and detection reactions is large compared to the 
neutrino mass (or compared to the mass differences). Note that if the condition 
opposite to $(i)$ is realized, the probabilities of flavour transitions also 
take a universal form, as in that case they simply correspond to averaged 
oscillations. 

(7) The stationary source approximation is valid when the time-dependent 
features of the neutrino emission and absorption processes are either 
absent or irrelevant, so that one essentially deals with steady neutrino 
fluxes. Integration over the neutrino detection time then results in the 
equivalence of the oscillation picture to that in the ``same energy'' 
approximation.

We have also demonstrated that, although calculations in the wave packet 
approach to neutrino oscillations are not manifestly covariant, they 
nevertheless result in a Lorentz invariant oscillation probability. 
This holds both in the case when the coherent production/detection and  
coherent propagation conditions are satisfied, so that the oscillation 
probability takes its standard form, and in the general case, when these 
conditions are not necessarily fulfilled.

\appendix
\renewcommand{\theequation}{\thesection\arabic{equation}}

\appsection
\renewcommand{\thesection}{\Alph{section}}
\section*{Appendix \Alph{section}: 
Dependence of $f_i^{S,D}(p)$ and $G_i(L-v_{gi}t)$ on the index $i$}
For ultra-relativistic neutrinos to an extremely good 
approximation
\be
f_i^{S,D}(p) = f^{S,D}(p, E_i(p))\,,
\label{eq:h}
\ee
i.e. the momentum distribution functions $f_i^{S,D}(p)$ depend on the index 
$i$ only through their dependence on the neutrino energy $E_i(p)=\sqrt{p^2
+m_i^2}$. The same applies to the shape factors $G_i(L-v_{gi}t)$. 
Indeed, these functions 
depend on $i$ through the neutrino mass dependence of the phase space volumes 
and of the amplitudes of the neutrino emission and detection processes. The 
phase space volumes depend on $m_i$ only through $E_i(p)$; the production and 
detection amplitudes can in principle depend directly on $m_i$, e.g., due to 
the chiral suppression, as in the case of $\pi^\pm$ or $K^\pm$ decays. However, 
the corresponding contributions to the total amplitudes are completely 
negligible compared to the main contributions, which in this case are 
proportional to the masses of charged leptons.  
 
Since $v_{gi}=\sqrt{E_i(p)^2-m_i^2}/E_i(p)$, eq.~(\ref{eq:h}), in particular, 
means that in the limit when $r=1$ (i.e. $v_{gi}=v_{gk}$) one has 
$f_i^{S,D}(p)=f_k^{S,D}(p)$ and $G_i=G_k$.

\appsection
\renewcommand{\thesection}{\Alph{section}}
\section*{Appendix \Alph{section}: 
Lorentz invariance of the oscillation probability and Gaussian wave packets}

We illustrate here how our general proof of Lorentz invariance of the 
oscillation probability, given in Sec.~\ref{sec:Lorentz}, works in the 
particular case of Gaussian wave packets. 
To this end, we shall prove the invariance of the integral 
\be
\tilde{I}_{ik}=\int dt\,G_i(L-v_{gi}t) G_k^*(L-v_{gk}t)\,.
\label{eq:tildeI}
\ee
Lorentz invariance of the oscillation probability (\ref{eq:P3}) will then 
follow from the definition of $I_{ik}$ in eq.~(\ref{eq:Iik1}) and Lorentz 
invariance of the phase difference $\Delta\phi$ (\ref{eq:phase1}). 

The normalized Gaussian wave packets of the neutrino mass eigenstates in 
the laboratory frame are \cite{Giunti:1997wq} 
\be
G_i(L-v_{gi}t)=\frac{v_{gi}^{1/2}}{(2\pi \sigma_{xi}^2)^{1/4}}
\,e^{-\frac{(L-v_{gi}t)^2}{4\sigma_{xi}^2}}\,.
\label{eq:Gauss1}
\ee
Here we have taken into account that the lengths $\sigma_{xi}$ of the wave 
packets corresponding to different mass eigenstates $\nu_i$ are in general 
different. Even though the dependence of $\sigma_{xi}$ on $i$ is expected to 
be extremely weak, we take it into account here, to be in line with the 
transformation law (\ref{eq:sigmax2}) which depends on the group velocity of 
the wave packet (one can neglect this dependence at the end of the 
calculation). Note that the exponential factor in (\ref{eq:Gauss1}) is Lorentz 
invariant (cf. eqs.~(\ref{eq:new}) and (\ref{eq:sigmax2})), whereas the 
pre-exponential factor is not.  

Consider the integral $\tilde{I}_{ik}$ with Gaussian wave packets 
$G_i(L-v_{gi}t)$ from (\ref{eq:Gauss1}): 
\be
\tilde{I}_{ik}=\frac{(v_{ki} v_{gk})^{1/2}}{(2\pi \sigma_{xi}\sigma_{xk})^{1/2}}
\,\int dt\, e^{-\frac{(L-v_{gi}t)^2}{4\sigma_{xi}^2}
-\frac{(L-v_{gk}t)^2}{4\sigma_{xk}^2}}\,.
\label{eq:tildeI2}
\ee
Direct calculation of the integral yields 
\be
\tilde{I}_{ik}=\left(\frac{2\, \sigma_{xi}\sigma_{xk}\, v_{gi}v_{gk}}
{\sigma_{xi}^2 v_{gk}^2+\sigma_{xk}^2 v_{gi}^2}\right)^{1/2}
e^{-\frac{L^2(v_{gi}-v_{gk})^2}{4(\sigma_{xi}^2 v_{gk}^2+\sigma_{xk}^2 
v_{gi}^2)}}\,\simeq\, 
e^{-\frac{L^2 \Delta v_{g}^2}{8\sigma_{x}^2 v_{g}^2}}\,.
\label{eq:tildeI3}
\ee
Here in the last equality we took into account that for relativistic or 
quasi-degenerate neutrinos one can set $\sigma_{xi}=\sigma_{xk}\equiv 
\sigma_x$ and also neglect the difference between $v_{gi}$ and $v_{gk}$ 
in the pre-exponential factor, while retaining the difference $\Delta v_g= 
v_{gi}-v_{gk}$ to leading order in the exponent, where it is multiplied by 
the macroscopic distance $L$. The resulting expression can be written as 
$\tilde{I}_{ik}\simeq e^{-(L/l_{\rm coh})^2}$, i.e. $\tilde{I}_{ik}$ is 
the overlap integral, which takes into account the suppression of neutrino 
oscillations when the baseline exceeds the coherence length and the wave 
packets corresponding to different mass eigenstates separate and cease to 
overlap. 

Let us now discuss the properties of $\tilde{I}_{ik}$ with respect to the 
Lorentz boosts. By changing the integration variable, eq.~(\ref{eq:tildeI2}) 
can be rewritten as 
\be
\tilde{I}_{ik}=\frac{(v_{gi} \sigma_{xk})^{1/2}}{(2\pi 
v_{gk}\sigma_{xi})^{1/2}}
\,\int \frac{ds}{\sigma_{xk}}\, 
e^{-\frac{[L(1-r)+s r]^2}{4\sigma_{xi}^2}-\frac{s^2}{4\sigma_{xk}^2}
}\,,
\label{eq:tildeI4}
\ee
where $s=L-v_{gk}t$.
As follows from eqs.~(\ref{eq:new}) and (\ref{eq:sigmax2}), the integration 
measure $ds/\sigma_{xk}$ is Lorentz invariant, and so is the exponential 
factor in the integrand (note that it coincides with 
that in eq.~(\ref{eq:tildeI2}) and has just been rewritten in terms of 
the new integration variable); therefore, the integral in (\ref{eq:tildeI4}) 
is Lorentz invariant. In a moving frame, the normalization factor of the 
Gaussian wave packet (\ref{eq:Gauss1}) has to be re-calculated (see the 
discussion below), which gives $\sigma_{x i,k}\to\sigma_{x i,k}'$, $v_{g i,k}
\to (v_{gi,k}'-u)$; taking this into account and making use of 
eqs.~(\ref{eq:relat}) and (\ref{eq:sigmax2}), it is easy to show that the 
factor in front of the integral in eq.~(\ref{eq:tildeI4}) is also invariant 
under Lorentz transformations. Thus, the quantity $\tilde{I}_{ik}$ is Lorentz 
invariant.

It is instructive to check explicitly how Lorentz invariance works in the case 
of Gaussian wave packets. Consider the overlap integral $\tilde{I}_{ik}'$ in a 
reference frame where the neutrino source and detector are boosted with a 
speed $u$ in the direction of motion of the neutrino. It should be stressed 
that one cannot obtain $\tilde{I}_{ik}'$ by simply replacing the unprimed 
quantities in eq.~(\ref{eq:tildeI3}) by the primed ones -- the integral has to 
be calculated anew. This is related to the important difference between the 
properties of  $L$ and $L'$. These are the distances neutrinos propagate, 
respectively, in the laboratory frame and in the moving frame, and they are 
defined as the differences between the corresponding coordinates of the 
neutrino absorption and emission points. However, while $L$ is a fixed 
distance between the neutrino source and detector in the laboratory 
frame, which does not depend on the neutrino propagation time $t$, the 
corresponding distance $L'$ in the moving frame is not fixed and actually 
depends on $t'$, as follows from eq.~(\ref{eq:trans1}). In particular, $L'$ is 
not just the usual Lorentz-contracted value of $L$ (equal to $L/\gamma_u$); 
the latter would correspond to the difference of the detector and source 
coordinates in the moving frame {\sl at the same time} $t'$, whereas $L'$, by 
the construction of the effective wave packets, is the difference of the 
detector coordinate at the time $t'$ and the source coordinate at $t'=0$. 

The dependence $L'(t')$ can be found by substituting $t$ from the second 
equality in (\ref{eq:trans1}) into the first one, which gives 
$L'(t')=L/\gamma_u+u t'$. Taking this into account and replacing in 
eq.~(\ref{eq:tildeI2}) all unprimed quantities by the primed ones 
gives, after a simple calculation, 
\be
\tilde{I}_{ik}'=
\frac{[2\, \sigma_{xi}'\sigma_{xk}'\, (v_{gi}'-u)(v_{gk}'-u)]^{1/2}}
{[\sigma_{xi}'^2 (v_{gk}'-u)^2+\sigma_{xk}'^2 (v_{gi}'-u)^2]^{1/2}}\;
e^{-\frac{L^2(v_{gi}'-v_{gk}')^2}{4\gamma_u^2
[\sigma_{xi}'^2 (v_{gk}'-u)^2+\sigma_{xk}'^2 (v_{gi}'-u)^2]}}\,.
\label{eq:tildeI5} 
\ee
(Note that the normalization factor for $G_i'(L'-v_{gi}'t')$ also had to 
be re-calculated, since the normalization of $G_i(L-v_{gi}t)$ in 
(\ref{eq:Gauss1}) was obtained for $t$-independent $L$). 
{}From eqs.~(\ref{eq:relat}) and (\ref{eq:sigmax2}) one finds  
\bea
[\sigma_{xi}'\, \sigma_{xk}'\, 
(v_{gi}'-u)(v_{gk}'-u)]^{1/2}&=&\frac{(\sigma_{xi}\sigma_{xk}\,v_{gi} 
v_{gk})^{1/2}}{\gamma_u^3(1+u v_{gi})(1+u v_{gk})}\,, 
\nonumber \\
v_{gi}'- v_{gk}'&=&\frac{v_{gi}-v_{gk}}{\gamma_u^2 (1+u 
v_{gi})(1+u v_{gk})}\,,
\nonumber\\ 
\,\sigma_{xi}'^2 (v_{gk}'-u)^2+\sigma_{xk}'^2(v_{gi}'-u)^2 &=&
\frac{\sigma_{xi}^2 v_{gk}^2+\sigma_{xk}^2 v_{gi}^2}{\gamma_u^6 
(1+u v_{gi})^2(1+u v_{gk})^2}\,. 
\eea
Substituting this into (\ref{eq:tildeI5}) and comparing the 
result with (\ref{eq:tildeI3}), one can readily make sure 
that  $\tilde{I}_{ik}'=\tilde{I}_{ik}$.


\end{document}